\documentclass[10pt,pra,aps,amsmath,amssymb,twocolumn,superscriptaddress,hidelinks]{revtex4-2}

\usepackage{usepackages}
\newcommand{\PSY}[1]{{\color{black} #1}}

\begin{document}

\title{Mixed-symmetry superconductivity and the energy gap}

\author{P. Senarath Yapa\,\orcidlink{0000-0002-5031-4695}}
\email{pramodh.senarath@uibk.ac.at}
\affiliation{
Universität Innsbruck, Fakultät für Mathematik, Informatik und Physik,
Institut für Experimentalphysik, 6020 Innsbruck, Austria
}
\affiliation{Department of Physics, University of Alberta, Edmonton, AB, Canada T6G~2E1}

\author{X. Guo\,\orcidlink{0009-0008-4789-7309}}
\affiliation{Department of Physics, University of Alberta, Edmonton, AB, Canada T6G~2E1}

\author{J. Maciejko\,\orcidlink{0000-0002-6946-1492}}
\affiliation{Department of Physics, University of Alberta, Edmonton, AB, Canada T6G~2E1}
\affiliation{Theoretical Physics Institute \& Quantum Horizons Alberta, University of Alberta, Edmonton, Alberta T6G 2E1, Canada}

\author{F. Marsiglio\,\orcidlink{0000-0003-0842-8645}}
\email{fm3@ualberta.ca}
\affiliation{Department of Physics, University of Alberta, Edmonton, AB, Canada T6G~2E1}
\affiliation{Theoretical Physics Institute \& Quantum Horizons Alberta, University of Alberta, Edmonton, Alberta T6G 2E1, Canada}

\begin{abstract}

The symmetry of the superconducting order parameter, or simply the ``gap'', provides certain constraints on the actual mechanism that gives rise to pairing and ultimately to superconductivity. In this work we show \PSY{how superconducting phases with mixed singlet-triplet symmetries can arise below $T_c$} for a generic tight-binding model. We first examine the 1D case to better illustrate the prevalence of symmetry-breaking transitions below $T_c$, and then the more realistic 2D case. In both cases we illustrate the implication for spectroscopic investigations of the energy gap by calculating the density of states for different temperatures below $T_c$. \PSY{We find that the structure of the density of states near $T_c$ can vary dramatically from its form near $T=0$}. A complete picture of the superconducting symmetry can only be attained if measurements are made over the entire temperature range.
\end{abstract}

\date{\today }
\maketitle

\section{Introduction}

The unambiguous experimental determination of the order parameter of superconducting materials remains a challenge~\cite{Norman2011}. Typically, the amplitude of the order parameter is measured by single-electron tunneling~\cite{fischer07} or in photoemission experiments~\cite{ARPESreview}; more ideally, if the phase is also measured, a better diagnostic of the symmetry can be performed \cite{vanHarlingen1995}. The symmetry of the order parameter remains an important diagnostic for the mechanism by which pairing arises. Within a couple of years of Bardeen-Cooper-Schrieffer (BCS) theory \cite{BCS1957} --- where initially only an $s$-wave order parameter was considered --- a number of researchers considered higher angular momentum pairing; some of these had in mind a superfluid phase in $^3$He, while others simply had a more general framework for superconductivity in materials (e.g. Pitaevskii~\cite{Pitaevskii1959} motivated by Landau, Brueckner et al. \cite{brueckner1960level}, Emery and Sessler \cite{emery1960possible} motivated by Mottelson, Thouless \cite{thouless1960perturbation}, Anderson and Morel \cite{anderson1961generalized}, Balian and Werthamer \cite{balian1963superconductivity}). Some of the more personal aspects of these developments, particularly in connection with superfluidity in $^3$He, are nicely described in a Special Issue of the Journal of Low Temperature Physics (Volume 164, Issue 3-4) with articles by Anderson~\cite{Anderson2011}, Sessler~\cite{Sessler2011}, Lee and Leggett ~\cite{Lee2011}, and Pitaevskii~\cite{Pitaevskii2011}.

It is important to realize that the symmetry is uniquely determined at the critical temperature $T_c$ of a superconducting phase. The emergent symmetry is the ``winner'' of a competition amongst the different symmetries, and these cannot mix at the transition due to the linearization of the gap equation. However, below $T_c$, the gap equation is non-linear, and in principle, nothing prevents the mixing of different order parameter symmetries. This is made abundantly clear in works by Annett~\cite{Annett1990}, Sigrist and Ueda~\cite{SigristUeda1991,Sigrist2005}, and \PSY{many subsequent publications~\cite{NORMAND1994,ODonovan1995,Musaelian1996,betouras1997,Mitra1998,ANGILELLA2001,BASU2006,Medvedev2012,Timirgazin2019,Akbar2024,Sutradhar2024,Kheirkhah2020}}. Moreover, concrete calculations have been made by S\"orensen et al.~\cite{Sorensen1991} \PSY{ and Kuboki~\cite{Kuboki2001}}, and more recently by Nayak and Kumar~\cite{Nayak_2018} and Hutchinson and Marsiglio~\cite{Hutchinson_2020}, all in the context of tight-binding models. 

The somewhat surprising outcome of these studies so far is that a mixed-symmetry order parameter at low temperatures is actually a {\it likely} outcome as a function of coupling parameters. Such a scenario has tremendous impact on the interpretation of experiments; while the aforementioned connection between order parameter symmetry and mechanism is applicable at the superconducting transition temperature, the order parameter symmetry is currently most easily measured in experiments at very low temperature, where the order parameter amplitude is a maximum. If the symmetry has undergone a change in the interim temperatures, then the connection to mechanism is not so clear.

In this study we address these problems and map out the phase boundaries for a tight-binding extended Hubbard model (eHM), using BCS theory for a translationally invariant system. We study this model in one dimension (1D) and two dimensions (2D) with nearest-neighbour interactions, allowing for the coexistence of both spin-singlet and spin-triplet order parameters. This means that for the 1D system only $s$-wave and $p$-wave superconducting order parameters are possible, while in the 2D system a $d$-wave order parameter is also allowed. We map out a phase diagram that illustrates the competition and mixing of order parameters with different symmetries. These calculations institute mean-field approximations, with the caveat that in many instances the same calculations have more validity than expected for a mean-field solution, as shown in various works \cite{richardson63,richardson64,richardson77,combescot13}.

The outline of the paper is as follows: in Section~\ref{SecI} we introduce the eHM, its mean-field approximation and the order parameters in the 1D and 2D systems. In Section~\ref{SecII} and Section~\ref{SecIII}, we consider the superconducting phases of the 1D lattice and 2D square lattices respectively. The 1D problem is considered first because it is simpler in many respects, and serves as a ``training ground'' for the 2D calculations. Both sections are organized in the following way. We first calculate the phase diagrams at the critical temperature for the first superconducting transition, $T_{c1}$. We plot the $T_{c1}$ phase diagrams for various $n_e$, and show how select points in the parameter space evolve towards mixed-symmetry solutions upon cooling towards $T=0$. We then show the full $T=0$ phase diagrams for the same values $n_e$. Finally, we calculate the density of states (DOS) of these superconducting phases, plotting the evolution of the DOS as a function of frequency when transitioning between pure and mixed-symmetry superconducting order by reducing the temperature. We focus on the DOS as it is a crucial diagnostic measured by tunneling for characterizing the symmetry of superconducting materials~\cite{Giaever1962,Tsuei2000,Hashimoto2014}, but clearly other experimental probes (e.g. photoemission, optical conductivity) will similarly be affected by a change in symmetry.

\section{The extended Hubbard Model and its mean-field solution}\label{SecI}
 Our starting point is the eHM with nearest-neighbour hopping $t$, chemical potential $\mu$ and  on-site interaction $U$, and nearest-neighbour interaction $V$,
\begin{equation}
\begin{aligned}
\hham = & -t \sum_{\langle i,j\rangle, \alpha} \left( \hat{c}_{i \alpha}^{\dagger} \hat{c}_{j \alpha} + \hat{c}_{j \alpha}^{\dagger} \hat{c}_{i \alpha} \right) -\mu \sum_{i, \sigma} \hat{n}_{i \sigma}\\
&+U \sum_i \hat{n}_{i \uparrow} \hat{n}_{i \downarrow} +V \sum_{\langle ij\rangle, \alpha \beta} \hat{n}_{i \alpha} \hat{n}_{j \beta},
\end{aligned}    
\end{equation}
where $\hat{c}_{i \alpha}^{\dagger}$ ($\hat{c}_{i \alpha}$) creates (annihilates) an electron of spin $\alpha$ at site $i$, and $\hat{n}_{i \alpha}$ is the number operator for electrons. Summations over single indices are over all sites of the lattice; the summation $\langle i,j\rangle$ implies summation over nearest neighbours only, and these are counted only once. Periodic boundary conditions are used.

Following Ref.~\cite{Hutchinson_2020}, we Fourier transform the eHM and perform a mean-field approximation in the pairing channel to obtain the following mean-field Hamiltonian:

\begin{equation}\label{EqHamMF}
\hham_{\mathrm{MF}} = \sum_{\boldsymbol{k}, \alpha} \xi_{\boldsymbol{k}} 
\hat{c}_{\boldsymbol{k} \alpha}^{\dagger} \hat{c}_{\boldsymbol{k} \alpha} 
 - \frac{1}{2} \sum_{\boldsymbol{k}, \alpha \beta} 
\Big[ \Delta_{\boldsymbol{k} \alpha \beta} \hat{c}_{\boldsymbol{k} \alpha}^{\dagger} 
\hat{c}_{-\boldsymbol{k} \beta}^{\dagger} 
+ \text{h.c.} \Big] -E_{c}.
\end{equation}
where $\xi_{\boldsymbol{k}}$ is the single-particle energy with
\begin{align}
\xi_k &\equiv -2t \cos{k} - \mu \quad \text{in 1D}, \\
\xi_{\boldsymbol{k}} &\equiv -2t (\cos{k_x} + \cos{k_y}) - \mu \quad \text{in 2D}.
\end{align}
For simplicity, we have set the lattice spacing $a \equiv 1$.
In Eq.~\ref{EqHamMF}, $\Delta_{\boldsymbol{k}\alpha\beta}$ is the gap parameter and $E_{c}$ is a constant  energy due to the product of mean fields. The gap parameter obeys the following self-consistent
equation:
\begin{equation}    \label{EqChap1GapEq}
\Delta_{\boldsymbol{k}\alpha\beta} = -\frac{1}{N} \sum_{\boldsymbol{k}^{\prime}, \alpha^{\prime}\beta^{\prime}} V_{\alpha \beta \alpha^\prime\beta^\prime}\left(\boldsymbol{k}, \boldsymbol{k}^{\prime}\right) \Delta_{\boldsymbol{k}^\prime\alpha^\prime\beta^\prime}g_{\boldsymbol{k}^\prime},
\end{equation}
where $V_{\alpha \beta \alpha^\prime\beta^\prime}\left(\boldsymbol{k}, \boldsymbol{k}^{\prime}\right)$ is the momentum-space interaction potential, which contains contributions from both the on-site and nearest-neighbour interactions, and $g_{{\bf k}^\prime}$ is defined below. 
We will work with a spin-balanced system ($n_\uparrow = n_\downarrow = \frac12 n_e$), and since there are no spin-flip interactions in the extended Hubbard model, this means that our spin-triplet order parameters are equivalent to each other under spin rotation symmetry. This allows us to simplify the problem by choosing the spin-pairing channel most convenient for computation. We choose the subspace with projection $m_s = 0$, i.e. $|S,m_S\rangle = |0,0\rangle$ and $|1,0\rangle$ with only $\uparrow\downarrow$ and $\downarrow\uparrow$ pairing, which allows us to access both spin-singlet and spin-triplet order parameters. This amounts to only considering pairing interactions for $\alpha\beta\beta^\prime\alpha^\prime = \uparrow\downarrow\downarrow\uparrow$ and $\downarrow\uparrow\uparrow\downarrow$, which leads to a gap parameter of the form:
\begin{equation}\label{EqGapParam}
\begin{aligned}
\Delta_{\b{k}\alpha\beta} 
&=\begin{pmatrix}
0 & \Delta^\text{(s)}_{\b{k}} + \Delta^\text{(t)}_{\b{k}}\\
-\Delta^\text{(s)}_{\b{k}} + \Delta^\text{(t)}_{\b{k}} & 0
\end{pmatrix},
\end{aligned}
\end{equation}
where $\Delta^\text{(s)}_{\b{k}}$ and $\Delta^\text{(t)}_{\b{k}}$ are the spin-singlet and spin-triplet gap parameter components respectively. Here we have retained the additional subscripts $\alpha \beta$ on the left for clarity, but the first (second) row corresponds to $\alpha = \uparrow$ ($\downarrow$) and similarly for $\beta$ and the columns.

For the 1D eHM, because of the tight-binding form of $V_{\alpha \beta \alpha^\prime\beta^\prime}\left(\boldsymbol{k}, \boldsymbol{k}^{\prime}\right)$, these components can be expanded in terms of three complex-valued order parameters, $\{\Delta_0,\, \Delta_{s^*},\, \Delta_{p}\}$:
\begin{equation}
\begin{aligned}
\Delta_{k}^{\text {(s)}} & =\Delta_0+\Delta_{s^*} \cos k, \\
\Delta_{k}^{\text {(t)}} & =\Delta_{p}\sin k.
\end{aligned}
\end{equation}

For the 2D eHM, the gap parameter components can be expanded in terms of five complex-valued order parameters, $\{\Delta_0,\, \Delta_{s^*},\, \Delta_{d_{x^2-y^2}},\, \Delta_{p_x},\, \Delta_{p_y}\}$:
\begin{equation}\label{Eq2DDeltaks}
\begin{aligned}
\Delta_{\bm{k}}^{\text {(s)}} & =\Delta_0+\Delta_{s^*} s_{\bm{k}}+\Delta_{d_{x^2-y^2}}d_{\bm{k}}, \\
\Delta_{\bm{k}}^{\text {(t)}} & =\Delta_{p_x} \sin k_x+\Delta_{p_y} \sin k_y,
\end{aligned}
\end{equation}
where
% \begin{equation}
% \Delta_{\boldsymbol{k}\alpha\beta} = -\frac{1}{N} \sum_{\boldsymbol{k}^{\prime} \alpha^{\prime}\beta^{\prime}} V_{\alpha \beta \alpha^\prime\beta^\prime}\left(\boldsymbol{k}, \boldsymbol{k}^{\prime}\right) \Big\langle \hat{c}_{-\boldsymbol{k}^{\prime} \alpha^\prime} \hat{c}_{\boldsymbol{k}^{\prime} \beta^{\prime}}\Big\rangle, \label{eq:GapDelkalphbeta}
% \end{equation}
\begin{align}
    s_{\boldsymbol{k}} &\equiv \frac{1}{2}\left(\cos k_x+\cos k_y\right),\label{eq:sk}\\
    d_{\boldsymbol{k}} &\equiv \frac{1}{2}\left(\cos k_x-\cos k_y\right).    \label{eq:dk}
\end{align}

\begin{figure*}[t]
    \includegraphics[width=0.85\linewidth]{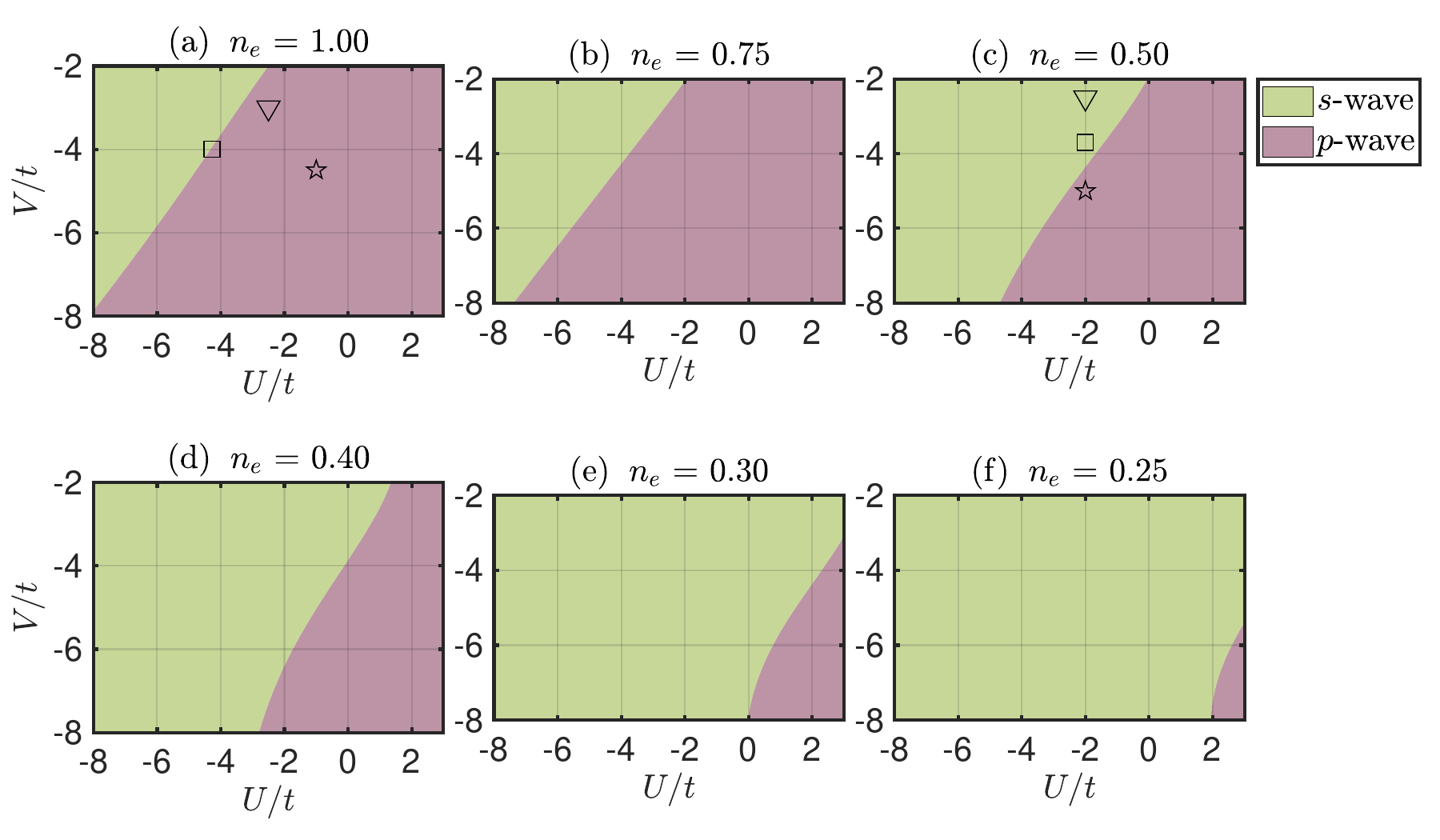}
        \captionsetup{justification=justified, singlelinecheck=false} % Justify multi-line captions
    \caption{\justifying % Ensure justification for the caption
     1D $T_{c1}$ phase diagrams for various $n_e$ using $N = 300$ lattice sites. The $\bigtriangledown$, $\square$ and {\small{\FiveStarOpen}} markers on subfigures (a) and (c) correspond to the values of $U$ and $V$ for which the order parameters are plotted in Fig.~\ref{Fig:1Dne100Cooling} and Fig.~\ref{Fig:1Dne050Cooling}.
     %, and later on the density of states in Fig.~\ref{Fig:1Dne100DOS} and Fig.~\ref{Fig:1Dne050DOS}.
     }
    \label{Fig1DBCSTcDiagrams}
\end{figure*}

\PSY{Note that for the 2D square lattice, the appropriate point group is $C_{4v}$, with the rotation axis along $\hat{z}$, perpendicular to the plane. The irreducible representations (irreps) of this group can be grouped into the 1D, parity-even irreps $\left(A_1, A_2, B_1, B_2\right)$, and the 2D, parity-odd irrep ($E$). The interaction, $V_{\alpha \beta \alpha^\prime\beta^\prime}\left(\boldsymbol{k}, \boldsymbol{k}^{\prime}\right)$, can be decomposed using basis functions belonging to each irrep. Thus we can identify the $\{\Delta_0,\, \Delta_{s^*}\}$ order parameters with the $A_1$ irrep, $ \Delta_{d_{x^2-y^2}}$ with the $B_1$ irrep and the $\{ \Delta_{p_x},\, \Delta_{p_y}\}$ order parameters with the $E$ irrep. These are the only possible order parameters in this model due to restricting ourselves to on-site and nearest-neighbour interactions. In principle, by introducing longer-range interactions, we could obtain a larger set of order parameters belonging to all the irreps of $C_{4v}$.}

Inserting these gap parameter components into Eq.~(\ref{EqChap1GapEq}), we obtain
self-consistent equations for each order parameter. We solve these coupled, non-linear equations for a chosen temperature, $T$, and electron density, $n_e$. We fix $n_e$ via the following equation:
\begin{equation}\label{EqNumberEq}
        n_e = 1 - \frac2N \sum_{\bm{k}} \xi_{\bm{k}}g_{\bm{k}},
\end{equation}
where  $g_{\boldsymbol{k}} \equiv \frac{1}{2 E_{\boldsymbol{k}}}\left[1-2 f\left(E_{\boldsymbol{k}}\right)\right]$, $f(E_{\bm{k}})$ is the Fermi-Dirac distribution and $ E_{\bm{k}}= \sqrt{\xi_{k}^2 +\left|\Delta_{\bm{k}}^{\text {(s)}}\right|^2+\left|\Delta_{\bm{k}}^{\text {(t)}}\right|^2}$. 

Note that the singlet and triplet order parameters in $E_{\boldsymbol{k}}$ add in quadrature due to the assumption that the gap parameter is unitary, i.e. 
\PSY{
\begin{equation}\label{EqUnitarity}
\Delta_{\b{k}} ^{\dagger} \Delta_{\b{k}} =\left|\Delta_{\b{k}} \right|^2 \mathbbm{1},
\end{equation}
where $\left|\Delta_{\b{k}} \right|^2\equiv$ $\frac{1}{2}  \Tr[\Delta_k^{\dagger} \Delta_k]$. }This unitarity condition is what allows us to write down a self-consistent equation for the gap parameter. A non-unitary gap parameter allows for the mean fields for spin $\uparrow\uparrow$ and $\downarrow\downarrow$ pairing to be arbitrary, which arises in spin-imbalanced systems or with interactions which favour certain spin orientations. Most superconducting systems --- both those investigated theoretically and measured experimentally --- have unitary gap parameters, and the mechanisms which stabilize non-unitary superconductivity are poorly understood~\cite{Ramires_2022}. 

Choosing the $m_s = 0$ pairing subspace guarantees the gap is unitary for a pure spin-triplet pairing. In the case of mixed-symmetry superconductivity, unitarity introduces an additional constraint. By inserting Equation~\ref{EqGapParam} into Equation~\ref{EqUnitarity}, we see that unitarity is only guaranteed if:
\PSY{\begin{equation}\label{EqUnitaryConstraint}
    \mathfrak{Re}\left[ \left(\Delta^\text{(s)}_{\b{k}}\right)^* \left(\Delta^\text{(t)}_{\b{k}}\right)\right] = 0,
\end{equation}}
where \PSY{$\mathfrak{Re}$ is the real} part of the expression in square brackets. This constraint is satisfied by the singlet and triplet order parameters having a $\pi/2$ phase difference between them when they co-exist. For example, an $s+d+ip$ phase is unitary, while an $s+d+p$ phase is non-unitary and therefore not considered in this study.

% \sout{\PSY{In the case of mixed-symmetry superconductivity, unitarity introduces an additional constraint. By inserting Equation~\ref{EqGapParam} into Equation~\ref{EqUnitarity}, we see that unitarity is only guaranteed if:}}
% \sout{
% \begin{equation}\label{EqUnitaryConstraint}
%     \mathfrak{Im}\left[ \left(\Delta^\text{(s)}_{\b{k}}\right)^* \left(\Delta^\text{(t)}_{\b{k}}\right)+\left(\Delta^\text{(s)}_{\b{k}} \right)\left(\Delta^\text{(t)}_{\b{k}}\right)^*\right] = 0,
% \end{equation}
% }
% \sout{\PSY{where $\mathfrak{Im}$ is the imaginary component of the expression in square brackets.  For example, an $s+d+ip$ phase is unitary, while an $s+d+p$ phase is non-unitary and therefore not considered in this study.}
% }}

Solving the self-consistent order parameter equations in tandem with Eq.~(\ref{EqNumberEq}), we obtain the set of order parameters which minimize the free energy. The mean-field free energy at a stationary point~\cite{Hutchinson_2020} is given by
\begin{equation}\label{EqfMFBCS}
\begin{split}
f_{\mathrm{MF}} = & \frac{1}{N} \sum_k\left(\xi_{\boldsymbol{k}} - E_{\boldsymbol{k}} + \left|\Delta_{\bm{k}}\right|^2 g_{\boldsymbol{k}}\right) \\
& + \frac{2 k_{\mathrm{B}} T}{N} \sum_{\boldsymbol{k}} \ln \left(1 - f\left(E_{\boldsymbol{k}}\right)\right) + \mu n_e.
\end{split}
\end{equation}
By choosing the set of solutions with the lowest free energy, we produce phase diagrams in the $U$-$V$ parameter space as a function of $n_e$ and $T$.

\section{1D Lattice} \label{SecII}
\subsection{1D phase diagrams at $T_{c1}$}
In 1D the order parameters $\{\Delta_0,\, \Delta_s^*,\, \Delta_{p}\}$  obey the self-consistent equations:
\begin{equation}\label{Eq1DCoupledEqs}
\begin{aligned}
&\Delta_0=-\frac{U}{N}\sum_{k} \left(\Delta_0+\Delta_{s^*} \cos{k}\right)g_k,\\
&\Delta_{s^*}=-\frac{2V}{N}\sum_{k}\left(\Delta_0+\Delta_{s^*} \cos{k}\right)g_k\cos{k},\\
&\Delta_p=-\frac{2V}{N}\sum_{k}\left(\Delta_p \sin^2{k}\right)g_k.
\end{aligned}
\end{equation}
Near $T_{c1}$ the order parameters are small, which allows us to linearize the equations about $\left|\Delta_k\right| = 0$, and set $E_k = |\xi_k|$. Thus $g_k$ is independent of the order parameters and the equations decouple into sets corresponding to the pure $s$-, $p$- and $d$-wave symmetry phases. In Fig.~\ref{Fig1DBCSTcDiagrams}  we plot the $T_{c1}$ phase diagrams for various $n_e$ --- we use these as a baseline for demonstrating how mixed-symmetry phases emerge upon further cooling.

We see that for $n_e = 1.00$, the $s$-wave region and the $p$-wave region are separated by $U\approx V$. At lower electron densities, the $s$-wave phase begins to dominate. Below $n_e \lessapprox 0.5$, we see that $s$-wave order is preferred even for repulsive $U$. Though this is primarily due to the $\Delta_{s^*}$ order parameter from the attractive $V$, the on-site $\Delta_0$ order parameter is also non-zero in this region at lower temperatures~\cite{PramodhPhDThesis}.

\begin{figure}[t]
    \centering
    \includegraphics[width=0.9\linewidth]{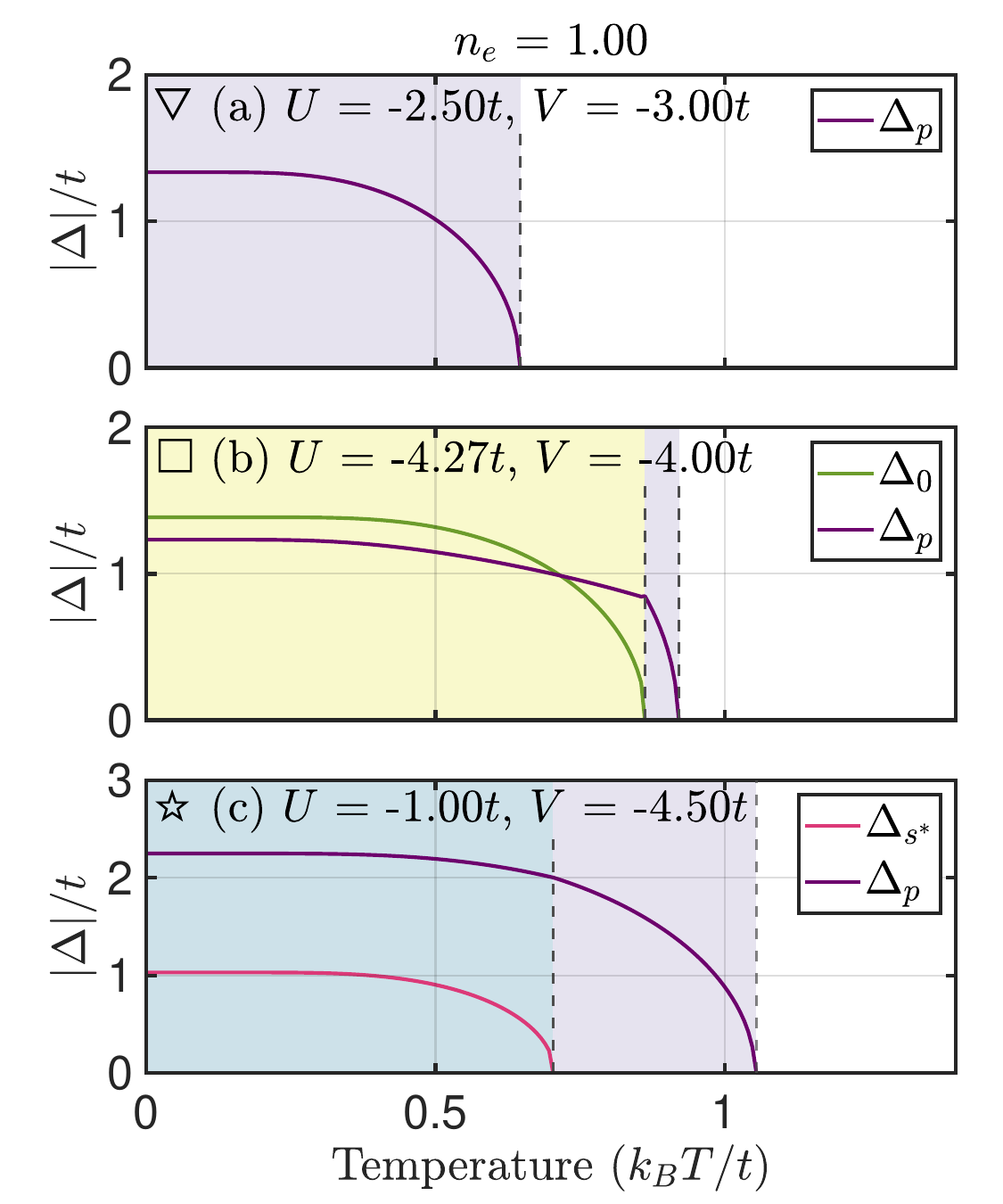}
    \caption{\justifying Phase transitions upon cooling the 1D $T_{c1}$ phase diagram for $n_e = 1.00$. The transition temperatures, $T_{ci}$, are indicated by vertical dashed gray lines, and the amplitude of the non-zero order parameters are plotted. The background colours correspond to the phase diagrams of Fig.~\ref{Fig1DBCSTcDiagrams} and Fig.~\ref{Fig1DBCST0Diagrams}.}
    \label{Fig:1Dne100Cooling}
\end{figure}

Below $T_{c1}$, the order parameters are coupled to each other and we must solve the full set of self-consistent equations. It is generically possible that several sets of order parameter solutions are stationary points of the free energy landscape. For example, both a pure $p$-wave phase ($\Delta^a_k = \Delta_{p}\sin k$) and a mixed $s+ip$ phase ($\Delta^b_k = |\Delta_0|+|\Delta_{s^*}| \cos k +i|\Delta_{p}|\sin k$) may be solutions of the 1D system, with the winner given by the phase with the lower free energy. \PSY{Note that the $\pi/2$ phase difference between the different order parameters (when they co-exist) can also be understood via a Ginzburg-Landau expansion of the free energy~\cite{Kuboki2001}.}

The free energies of a given solution vary as a function of $U$, $V$, $T$ and $n_e$; intersection of $f^a_{\mathrm{MF}}$ and $f^b_{\mathrm{MF}}$ then correspond to phase transitions between the solutions $\Delta^a_k$ and $\Delta^b_k$. To show these phase transitions, we select some points on the $U$-$V$ phase diagrams at $T_{c1}$ and plot the order parameters as a function of $T$. We show that multiple superconducting phase transitions can occur as we cool to $T=0$, and that for many of the $U$-$V$ points, the pure-symmetry phases at $T_{c1}$ give way to mixed-symmetry superconducting phases at low temperatures.

\subsection{1D mixed-symmetry phases below $T_{c1}$}

In Fig.~\ref{Fig:1Dne100Cooling}, we plot the absolute value of the non-zero order parameters as a function of temperature, for selected points ($\bigtriangledown$, $\square$ and {\small{\FiveStarOpen}}) on the $T_{c1}$ phase diagrams at half filling ($n_e = 1.00$). At $T_{c1}$, all three of the $U$-$V$ points correspond to the $p$-wave phase. 
% Note that the point (b) was especially chosen near the boundary of the $s$- and $p$-wave phases as it corresponds to a unique phase at $T=0$. \redd{What does this mean? --- it looks like a mixed phase. Unique in that it is on-site s + ip?}

\begin{figure}[t]
    \centering
    \includegraphics[width=0.9\linewidth]{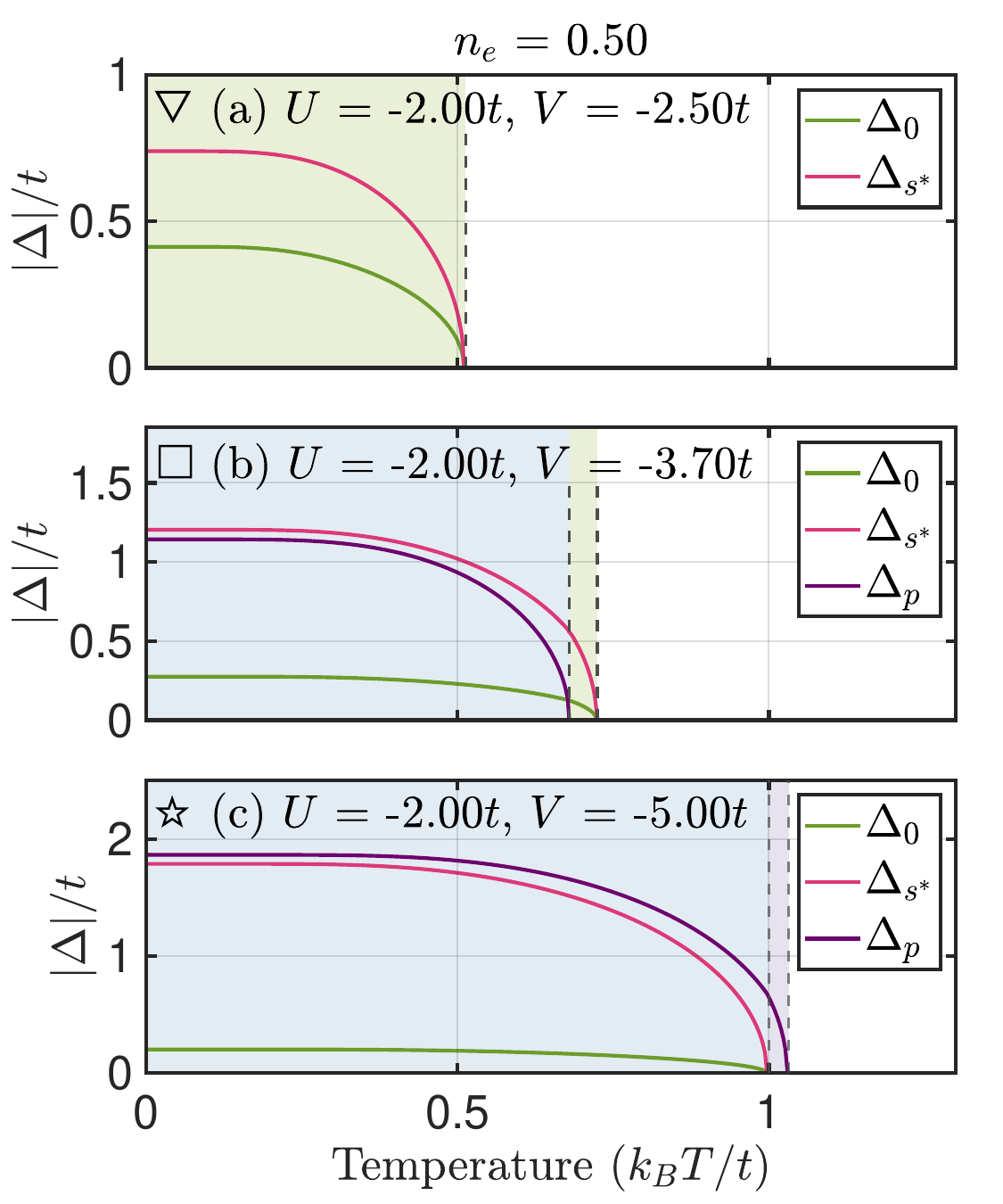}
    \caption{\justifying Phase transitions from cooling the 1D $T_{c1}$ phase diagram for $n_e = 0.50$. The transition temperatures, $T_{ci}$, are indicated by vertical dashed gray lines, and the amplitude of the non-zero order parameters are plotted. The background colours correspond to the phase diagrams of Fig.~\ref{Fig1DBCSTcDiagrams} and Fig.~\ref{Fig1DBCST0Diagrams}.}
    % \caption[Phase transitions from cooling the 1D $T_{c1}$ phase diagram for $n_e = 0.50$.]{Phase transitions from cooling (a) the 1D $T_{c1}$ phase diagram for $n_e = 0.50$. The transition temperatures are indicated by vertical dashed gray lines, and the non-zero order parameters are plotted  for a) $U/t = -2.00$, $V/t = -2.50$, b) $U/t = -2.00$, $V/t = -3.70$, and c) $U/t = -2.00$, $V/t = -5.00$.}
    \label{Fig:1Dne050Cooling}
\end{figure}

\begin{figure*}[ht]
    \includegraphics[width=0.95\linewidth]{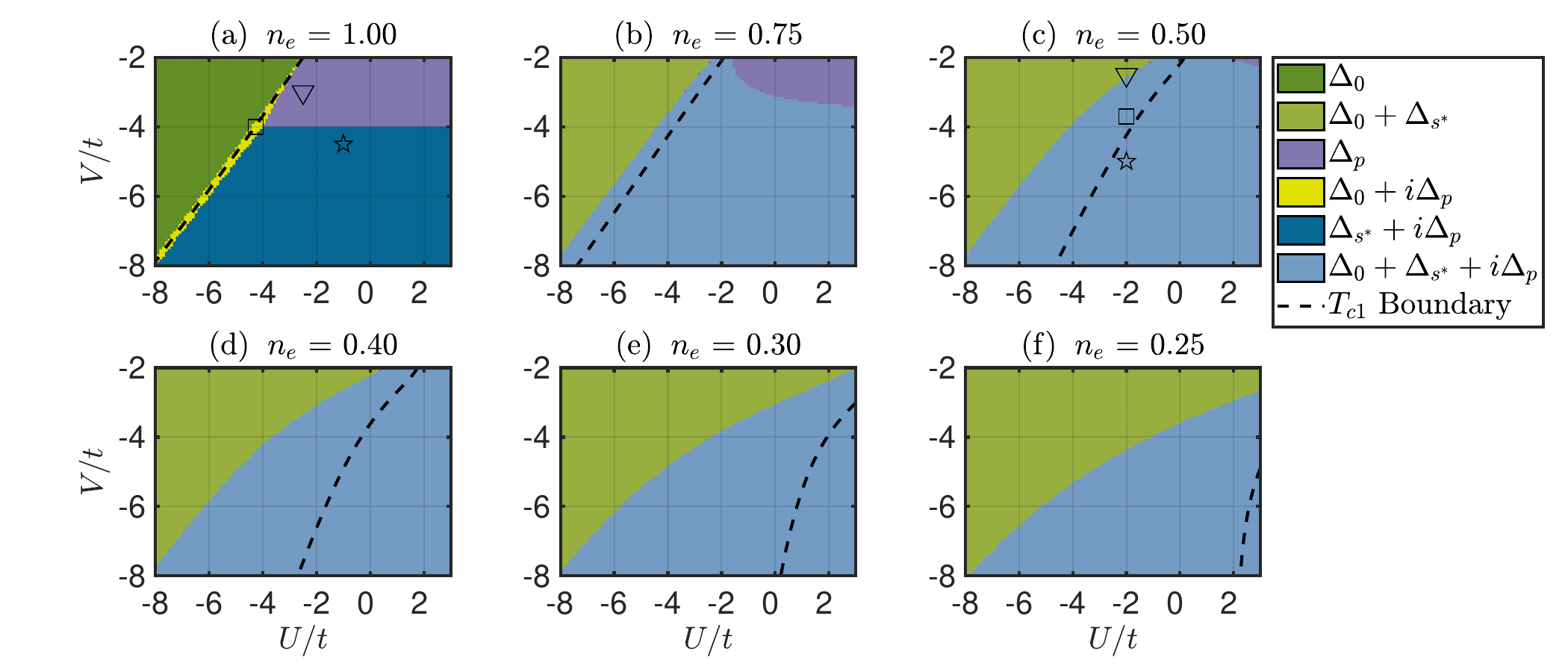}
        \captionsetup{justification=justified, singlelinecheck=false} % Justify multi-line captions
    \caption{\justifying % Ensure justification for the caption
   1D phase diagrams at $T = 0$ for various $n_e$ using $N = 500$ lattice sites. The $\bigtriangledown$, $\square$ and {\small{\FiveStarOpen}} markers on subfigures (a) and (c) correspond to the values of $U$ and $V$ for which the order parameters are plotted in Fig.~\ref{Fig:1Dne100Cooling} and Fig.~\ref{Fig:1Dne050Cooling}.
   %, and the density of states in Fig.~\ref{Fig:1Dne100DOS} and Fig.~\ref{Fig:1Dne050DOS}. \redd{again, address the ordering of these figures}
   }
    \label{Fig1DBCST0Diagrams}
\end{figure*}

In Fig.~\ref{Fig:1Dne100Cooling}(a), we see one superconducting transition into the $p$-wave phase at $T_{c1} = 0.647t/k_B$ which persists to $T = 0$. In Fig.~\ref{Fig:1Dne100Cooling}(b) we see the first superconducting transition into the $p$-wave phase occurs at $T_{c1} = 0.922t/k_B$. This is followed by a second transition at $T_{c2} = 0.862t/k_B$, into a mixed-symmetry phase of $p$- and on-site $s$-wave order. In Fig.~\ref{Fig:1Dne100Cooling}(c) the first superconducting transition occurs at $T_{c1} = 1.055t/k_B$, followed by a second transition at $T_{c2} = 0.704t/k_B$, this time into a mixed-symmetry phase of $p$-  and extended $s$-wave order.

Comparing Fig.~\ref{Fig:1Dne100Cooling}(b) and Fig.~\ref{Fig:1Dne100Cooling}(c), we see that the on-site $s$- and extended $s$-wave order parameters do not appear together even though they share the same symmetry. This is a general feature of half-filling for all $T$~\cite{PramodhPhDThesis}, and we will see that it also applies to the 2D case in the next section. For all other electron densities, the on-site $s$- and extended $s$-wave orders appear together.

In Fig.~\ref{Fig:1Dne050Cooling}, we show the behaviour of the order parameters upon cooling at quarter filling ($n_e = 0.50$). At $T_{c1}$, Fig.~\ref{Fig:1Dne050Cooling}(a) and Fig.~\ref{Fig:1Dne050Cooling}(b) correspond to an $s$-wave phase and Fig.~\ref{Fig:1Dne050Cooling}(c) corresponds to a $p$-wave phase. In Fig.~\ref{Fig:1Dne050Cooling}(a) we see the superconducting transition occurs at $T_{c1} = 0.513t/k_B$ into an $s$-wave phase consisting of both on-site $s$- and extended $s$-wave order. In this case, we see that there is no further superconducting transition, and this phase persists to $T = 0$. In Fig.~\ref{Fig:1Dne050Cooling}(b) the first superconducting transition is again an $s$-wave phase of both on-site $s$- and extended $s$-wave order, and occurs at $T_{c1} = 0.724t/k_B$. This is followed by a second transition at $T_{c2} = 0.679t/k_B$ into a fully mixed-symmetry phase. In Fig.~\ref{Fig:1Dne050Cooling}(c) the first superconducting transition occurs at $T_{c1} = 1.032t/k_B$ into a $p$-wave phase, followed by a second transition at $T_{c2} = 1.000t/k_B$, into the mixed-symmetry phase.

\subsection{1D phase diagrams at $T = 0$}

Having seen how select points on the $T_{c1}$ phase diagrams evolve upon cooling, we now set $T=0$ and show the ground-state phase diagrams in Fig.~\ref{Fig1DBCST0Diagrams}.

For the $T=0$ phase diagrams in Fig.~\ref{Fig1DBCST0Diagrams}, we plot the phase boundaries at $T_{c1}$ from Fig.~\ref{Fig1DBCSTcDiagrams} as a black dotted line for each $n_e$. The colours for the ground-states correspond to different combinations of $\{\Delta_0,\, \Delta_s^*,\, \Delta_{p}\}$ with fixed relative phases. Organized in this manner, we see that there are six different phases as indicated in the legend of Fig.~\ref{Fig1DBCST0Diagrams}. However, we note that the $\Delta_0$ phase and the $\Delta_0+ \Delta_{s^*}$ phase correspond to the same $s$-wave symmetry, and are not technically distinct phases. Similarly, the $\Delta_0 +i\Delta_p$ phase, $\Delta_{s^*}+i\Delta_p$ phase and the $\Delta_0 + \Delta_{s^*}+i\Delta_p$ have the same mixed $s+ip$ symmetry. We have chosen this colour scheme to highlight the fact that the two $s$-wave order parameters do not mix at $n_e = 1.00$.

We see that the $\Delta_0 +i\Delta_p$ phase only shows up at $n_e = 1.00$, in a narrow sliver under the black dotted $T_{c1}$ line. Away from half filling, we only have three distinct phases: (i) the pure $s$ ($\Delta_0+ \Delta_{s^*}$), (ii) the pure $p$ ($\Delta_p$), and (iii) the mixed-symmetry $s+ip$ ($\Delta_0 + \Delta_{s^*}+i\Delta_p$) phases. We also observe that at $n_e = 1.00$ the $T_{c1}$ line roughly coincides with a $T=0$ phase boundary (separating the pure on-site $s$-wave region from the other regions), but this coincidence does not occur away from half filling.

\subsection{1D density of states}

We can calculate the density of states (DOS) in the superconducting phases by

\begin{equation}
g(\omega)=\frac{1}{N} \sum_k\left[u_k^2 \delta\left(\omega-E_k\right)+v_k^2 \delta\left(\omega+E_k\right)\right],
\end{equation}

where

\begin{equation}
\begin{aligned}
& u_k^2=\frac{1}{2}\left(1+\frac{\xi_{k}}{E_k}\right), \\
& v_k^2=\frac{1}{2}\left(1-\frac{\xi_{k}}{E_k}\right).
\end{aligned}
\end{equation}

\begin{figure}[b]
    \centering
    \includegraphics[width=0.9\linewidth]{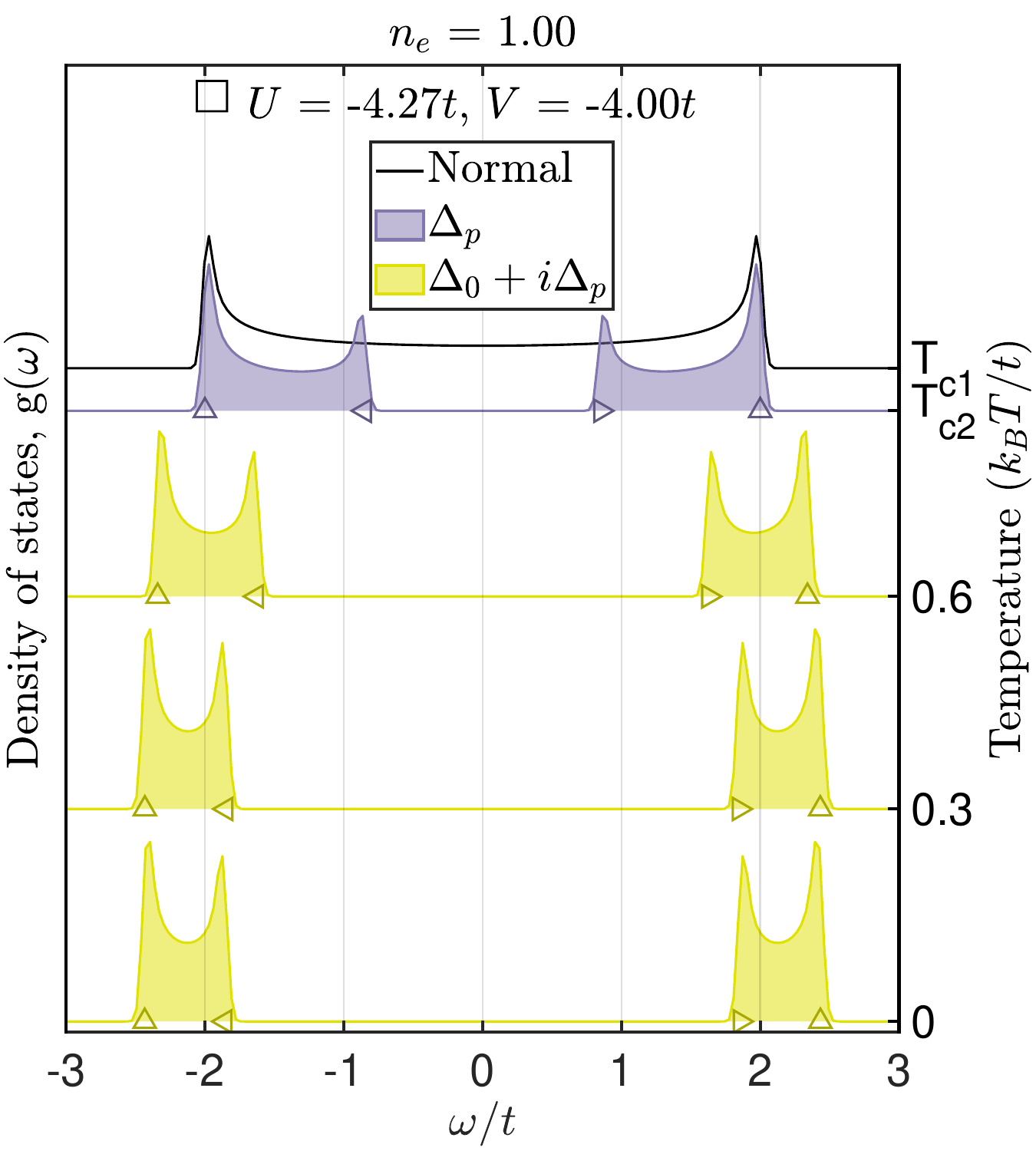}
    \caption{\justifying 1D density of states for $n_e = 1.00$ at $U/t = -4.27$ and $V/t = -4.00$ ($\square$ in subfigure (a) of Fig.~\ref{Fig1DBCSTcDiagrams} and Fig.~\ref{Fig1DBCST0Diagrams}). The transition temperatures are $T_{c1} = 0.922t/k_B$ and $T_{c2} = 0.862t/k_B$, and the colour of each DOS plot corresponds to its phase in its phase diagram. ``Normal'' in the legend refers to the normal state (non-superconducting) 1D density of states. The $\scriptstyle\triangle$ markers indicate the location of the band-edge peaks, and $\triangleright$ and $\triangleleft$ markers indicate the location of gap-edge DOS peaks.
    % At $T_{c1}$ the superconducting phase is a pure $d$-wave phase, and undergoes a phase transition to a mixed-symmetry $s+d+ip$ phase at $T_{c2} = 0.525t/k_B$.
    }
    \label{Fig:Waterfall1Dne100DOS}
\end{figure}

In our numerical implementation, we approximate the Dirac delta, $\delta(\omega)$, by a Gaussian:
\begin{equation}
\delta(\omega) \approx \frac{1}{\sqrt{2\pi \nu^2}} \exp\left(-\frac{\omega^2}{2\nu^2}\right),
\end{equation}
where $\nu$ is a broadening factor that determines the width of the Gaussian distribution; we choose $\nu/t = 0.03$ for all subsequent calculations. In the normal state, where $E_k = |\xi_{k}|$, the 1D DOS has a simple analytic form:

\begin{equation}
    g(\omega) = \frac{1}{\pi} 
%    \frac{1}
    \frac{\theta(2t - |\omega + \mu|)}
    {\sqrt{(2t)^2 -(\omega+\mu)^2}}.
%    {\sqrt{(2t+\mu)^2 - \omega^2}}.
\end{equation}
From this expression, we can observe that the 1D DOS has square root divergent Van Hove singularities at the band edges, i.e. at $\omega =  - \mu \pm 2t$. To show how this DOS evolves upon multiple transitions into different superconducting phases, we focus on two $U$-$V$ points: $U/t = -4.27$ and $V/t = -4.00$ for $n_e = 1.00$, and $U/t = -2.00$ and $V/t = -3.70$ for $n_e = 0.50$. These points are the $\square$ markers in subfigures (a) and (c) of Fig.~\ref{Fig1DBCSTcDiagrams} and Fig.~\ref{Fig1DBCST0Diagrams} respectively. We show the DOS for all the $\bigtriangledown$, $\square$ and {\small{\FiveStarOpen}} markers in Appendix~\ref{App1}.

In Fig.~\ref{Fig:Waterfall1Dne100DOS} at $n_e = 1.00$, the system transitions first into a pure $p$-wave phase and then into a mixed-symmetry phase of onsite $s$-wave and $p$-wave order. Just below $T_{c1}$, we see the opening of an energy gap of width $2|\Delta_p|$, leading to two new peaks (the BCS coherence peaks) at the gap edge. These appear due to the aggregation of states which were formerly within the gap. For the superconducting phases, we label the locations of the two band edge peaks by $\scriptstyle \triangle$ markers, and two gap edge peaks by the $\triangleright$ and $\triangleleft$ markers. Cooling below $T_{c2}$ into the mixed-symmetry $\Delta_0+i\Delta_p$ phase, we see that the gap widens but no new peaks appear in the DOS. 

\begin{figure}[b]
    \centering
    \includegraphics[width=0.9\linewidth]{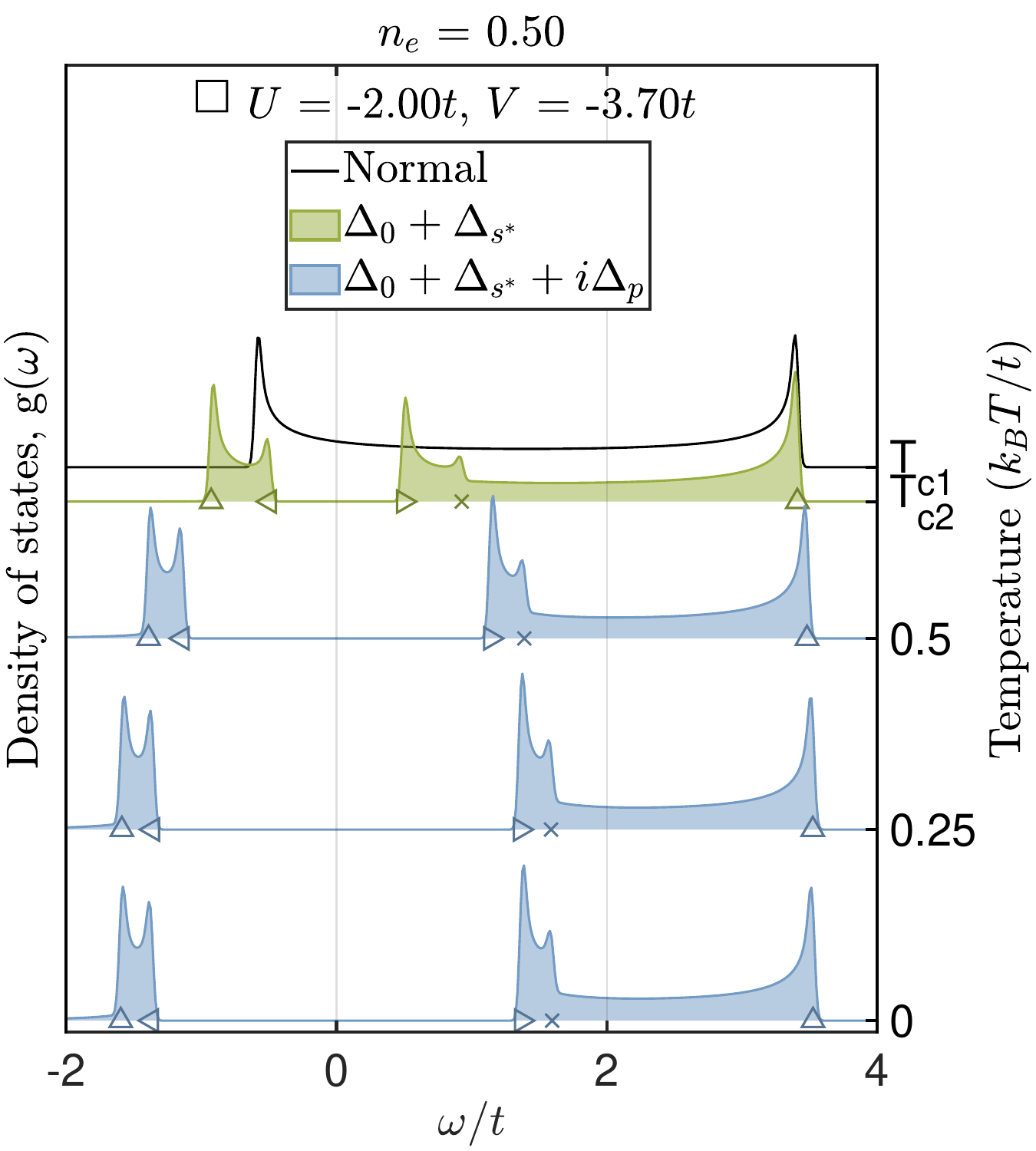}
    \caption{\justifying 1D density of states for $n_e = 0.50$ at $U/t = -2.00$ and $V/t = -3.70$ ($\square$ in subfigure (c) of Fig.~\ref{Fig1DBCSTcDiagrams} and Fig.~\ref{Fig1DBCST0Diagrams}).  The transition temperatures are $T_{c1} = 0.724t/k_B$ and $T_{c2} = 0.679t/k_B$, and the colour of each DOS plot corresponds to its phase in its phase diagram. ``Normal'' in the legend refers to the normal state (non-superconducting) 1D density of states.  The $\scriptstyle\triangle$ markers indicate the location of the band-edge peaks, the $\triangleright$ and $\triangleleft$ markers indicate the location of gap-edge DOS peaks, and the $\times$ markers indicate additional emergent peaks.
    % At $T_{c1}$ the superconducting phase is a pure $d$-wave phase, and undergoes a phase transition to a mixed-symmetry $s+d+ip$ phase at $T_{c2} = 0.375t/k_B$, and finally to a chiral $p_x+ip_y$ phase at $T_{c3} = 0.285t/k_B$
    }
    \label{Fig:Waterfall1Dne050DOS}
\end{figure}

\begin{figure*}[t]
    \includegraphics[width=0.85\linewidth]{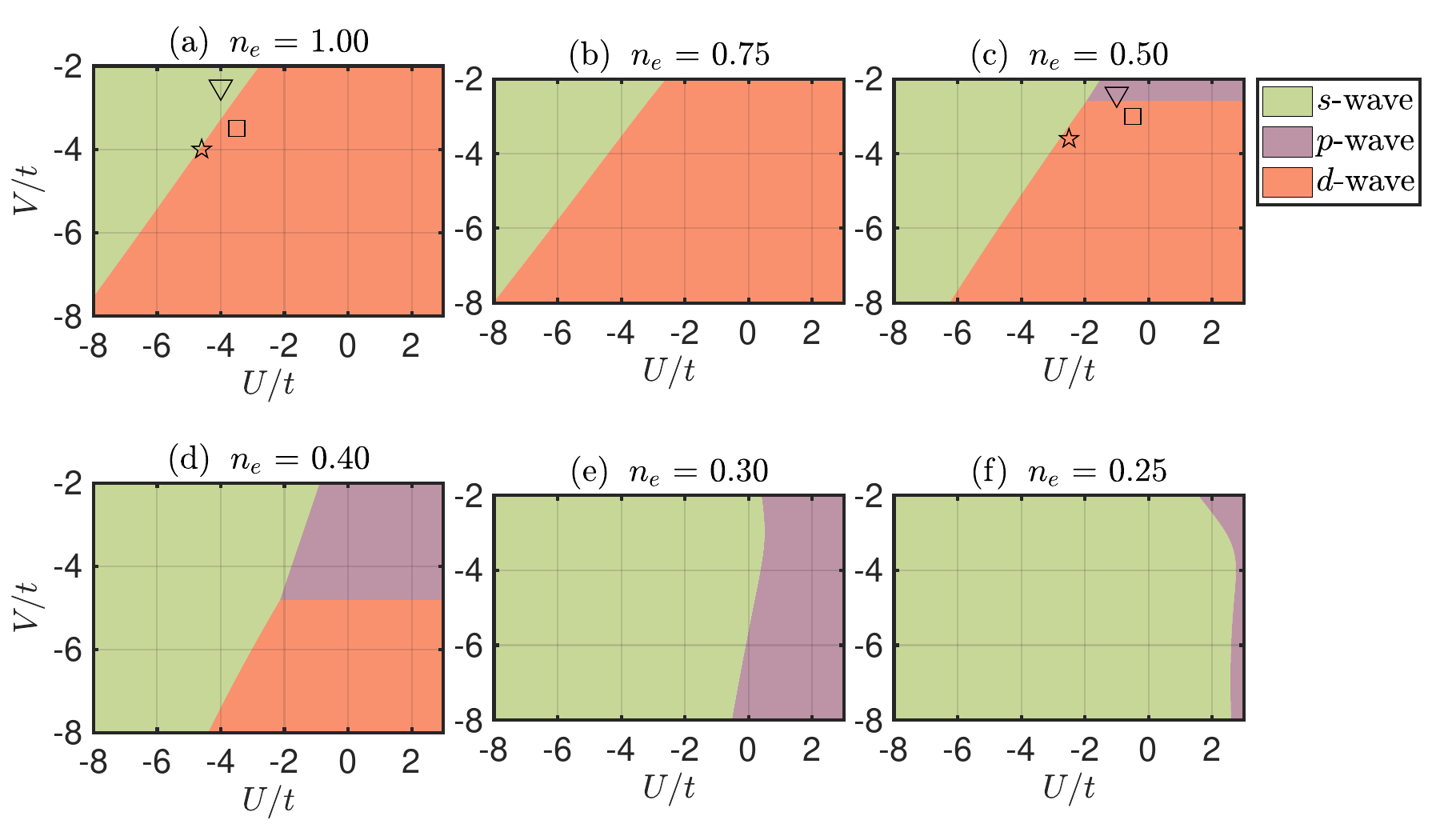}
        \captionsetup{justification=justified, singlelinecheck=false} % Justify multi-line captions
    \caption{\justifying % Ensure justification for the caption
    2D phase diagrams at $T_{c1}$ for various $n_e$ using $[N_x,\,N_y]$ = $[100,\,100]$ lattice sites. The $\bigtriangledown$, $\square$ and {\small{\FiveStarOpen}} markers on subfigures (a) and (c) correspond to the values of $U$ and $V$ for which the order parameters are plotted in Fig.~\ref{Fig:2Dne100Cooling} and Fig.~\ref{Fig:2Dne050Cooling}.}
    \label{Fig2DBCSTcDiagrams}
\end{figure*}

As noted in the preceding section, the two $s$-wave order parameters are allowed to mix below half-filling, which leads to new features in the DOS. In Fig.~\ref{Fig:Waterfall1Dne050DOS} at $n_e = 0.50$, the system transitions first into a pure $s$-wave phase and then into a mixed-symmetry phase of $s$-wave and $p$-wave order. In both these phases, an additional peak emerges in the positive frequency band of the DOS, labeled by the $\times$ marker in each plot. This peak occurs due to the stationary point at $k=0$ in the energy $E_{k}$. From this condition, we can derive that the peak occurs at

\begin{equation}\label{Eq1DvanHoveOmega}
    \omega_\times = \sqrt{(2t+\mu)^2+|\Delta_0+\Delta_{s^*}|^2},
\end{equation}
and thus originates from the two $s$-wave order parameters. We note that this peak coincides with the Van Hove singularity in the negative frequency band, i.e. the $\omega<0$ band edge peak occurs at $\omega^-_\triangle = -\omega_\times =-\sqrt{(2t+\mu)^2+|\Delta_0+\Delta_{s^*}|^2}$. We also note that the DOS is not strictly zero below $\omega^-_\triangle$, but has a ``soft shoulder''; this is most visible in the $T=0$ mixed-symmetry phase DOS. This shoulder terminates at $\omega = -\omega^+_\triangle$ with a small peak in the DOS; this is off-scale in Fig.~\ref{Fig:Waterfall1Dne050DOS}, but can be seen in Fig.~\ref{Fig:1Dne050DOS} of Appendix~\ref{App1}.

To summarize the superconducting DOS in 1D, we observe that the appearance of an additional order parameter away from half-filling (in this case, the  $\Delta_{s^*}$) leads to a new peak in the DOS. In the 1D system, the origin of this peak is not due to mixed-symmetry superconductivity, as Equation~\ref{Eq1DvanHoveOmega} only contains the two $s$-wave order parameters. However in the following section, we will demonstrate that in the 2D square lattice, the emergence of an additional peak in the DOS originates from mixed-symmetry superconducting order.

\section{2D square lattice}\label{SecIII}
\subsection{2D phase diagrams at $T_{c1}$}
We now proceed with the same analysis for the superconducting phases on the 2D square lattice. The five order parameters in 2D, $\{\Delta_0,\, \Delta_{s^*},\, \Delta_{d_{x^2-y^2}},\, \Delta_{p_x},\, \Delta_{p_y}\}$, obey the following self-consistent equations:
\begin{equation}\label{Eq2DOPEqs}
\begin{aligned}
& \Delta_0=-\frac{U}{N} \sum_{\bm{k}} \left(\Delta_0+\Delta_{s^*} s_{\bm{k}}+\Delta_{d_{x^2-y^2}} d_{\bm{k}}\right)g_{\bm{k}}, \\
& \Delta_{s^*}=-\frac{4 V}{N} \sum_{\bm{k}}  \left(\Delta_0+\Delta_{s^*} s_{\bm{k}}+\Delta_{d_{x^2-y^2}} d_{\bm{k}}\right)g_{\bm{k}}s_{\bm{k}}, \\
& \Delta_{d_{x^2-y^2}}=-\frac{4 V}{N} \sum_{\bm{k}}  \left(\Delta_0+\Delta_{s^*} s_{\bm{k}}+\Delta_{d_{x^2-y^2}} d_{\bm{k}}\right)g_{\bm{k}}d_{\bm{k}}, \\
& \Delta_{p_x}=-\frac{2 V}{N} \sum_{\bm{k}} \left(\Delta_{p_x} \sin k_x+\Delta_{p_y} \sin k_y\right) g_{\bm{k}}\sin k_x,\\
& \Delta_{p_y}=-\frac{2 V}{N} \sum_{\bm{k}} \left(\Delta_{p_x} \sin k_x+\Delta_{p_y} \sin k_y\right) g_{\bm{k}}\sin k_y.
\end{aligned}
\end{equation}
Solving these, we plot the $T_{c1}$ phase diagrams for various $n_e$ on the 2D square lattice in Fig.~\ref{Fig2DBCSTcDiagrams}.

We see that the $d$-wave order is dominant for $n_e = 1.00$ in Fig.~\ref{Fig2DBCSTcDiagrams}(a), but becomes increasingly more encroached upon as we lower the electron density. In these phase diagrams, $p$-wave order appears for first time in Fig.~\ref{Fig2DBCSTcDiagrams}(c) at $n_e = 0.50$, and becomes preferred over the $d$-wave order as $n_e$ is lowered. However the $s$-wave order becomes dominant over both $d$- and $p$-wave order for lower electron densities. This is true even for repulsive $U$ as seen in the $n_e=0.25$ phase diagram of Fig.~\ref{Fig2DBCSTcDiagrams}(f); though this $s$-wave phase is primarily composed of the $\Delta_{s^*}$ order parameter from the attractive $V$, the onsite $\Delta_0$ order parameter is also non-zero in this phase despite $U>0$~\cite{PramodhPhDThesis}.

\subsection{2D mixed-symmetry phases below $T_{c1}$}

We now proceed in the same manner as the previous section, and investigate how three points (marked by $\bigtriangledown$, $\square$ and {\small{\FiveStarOpen}}) on the $n_e = 1.00$ and $n_e = 0.50$ phase diagrams of Fig.~\ref{Fig2DBCSTcDiagrams} change upon cooling.

\begin{figure}[t]
    \centering
    \includegraphics[width=0.9\columnwidth]{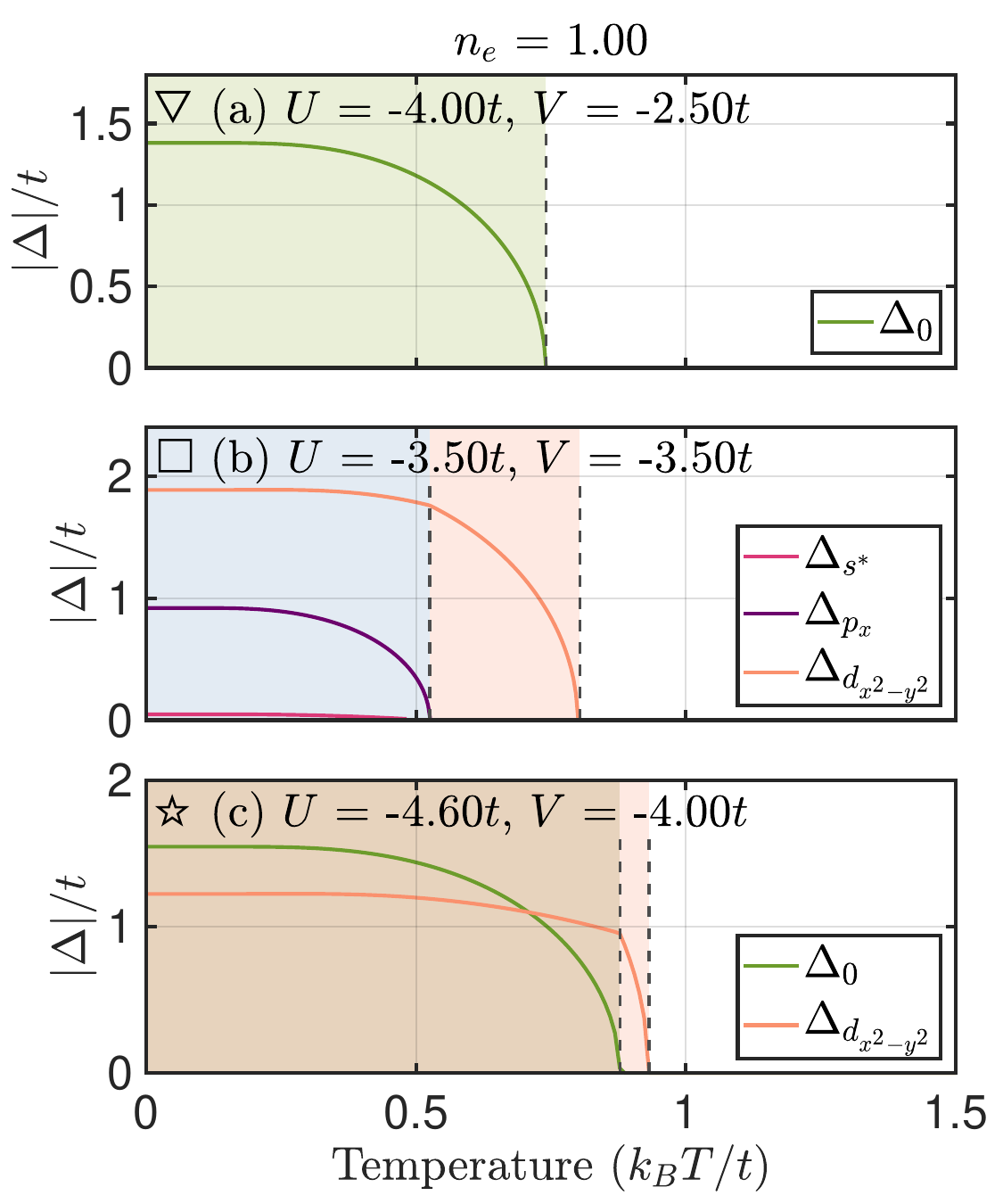}
    \caption{\justifying Phase transitions upon cooling the 2D $T_{c1}$ phase diagram for $n_e = 1.00$. The transition temperatures, $T_{ci}$, are indicated by vertical dashed gray lines, and the amplitude of the non-zero order parameters are plotted. The background colours correspond to the phase diagrams of Fig.~\ref{Fig2DBCSTcDiagrams} and Fig.~\ref{Fig2DBCST0Diagrams}.}
    \label{Fig:2Dne100Cooling}
\end{figure}

In Fig.~\ref{Fig:2Dne100Cooling}(a) we see one superconducting transition at $T_{c1} = 0.740t/k_B$ into a pure on-site $s$-wave phase, which persists to $T = 0$. In Fig.~\ref{Fig:2Dne100Cooling}(b) we see the first superconducting transition occur at $T_{c1} = 0.803t/k_B$, followed by a second transition at $T_{c2} =0.525t/k_B$ into a mixed-symmetry phase of extended $s$-, $d$- and $p$-wave order which we call the $s+d+ip$ phase. Note that the $s+d+ip$ phase is degenerate with the $s+d-ip$ phase by time-reversal symmetry. In general, all the phases with a complex order parameter of the form $A+iB$ are degenerate with $A-iB$, and thus spontaneously break time-reversal symmetry. We also do not specify the $p$-wave component of the $s+d+ip$ phase as the $s+d\pm ip_x$ phases are degenerate with the $s-d\pm ip_y$ phases; note that the $d$-wave order parameter acquires a minus sign in the latter phase, as they are related by the interchange of momenta $k_x \leftrightarrow k_y$.
\begin{figure}[t]
    \centering
    \includegraphics[width=0.9\linewidth]{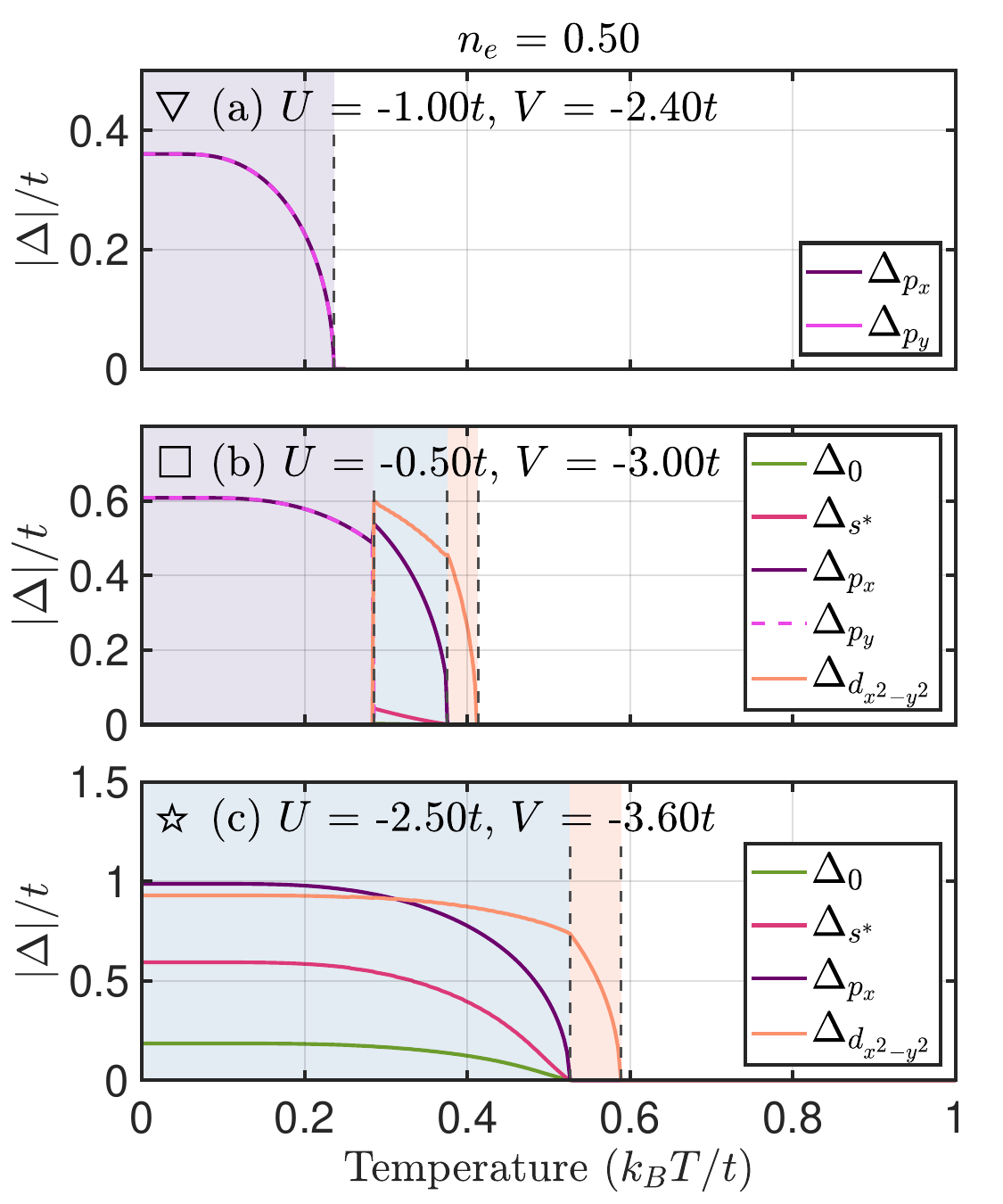}
    \caption{\justifying Phase transitions from cooling the 2D $T_{c1}$ phase diagram for $n_e = 0.50$. The transition temperatures, $T_{ci}$, are indicated by vertical dashed gray lines, and the amplitude of the non-zero order parameters are plotted. The background colours correspond to the phase diagrams of Fig.~\ref{Fig2DBCSTcDiagrams} and Fig.~\ref{Fig2DBCST0Diagrams}.}
    \label{Fig:2Dne050Cooling}
\end{figure}

\begin{figure*}[t]
    \includegraphics[width=1\linewidth]{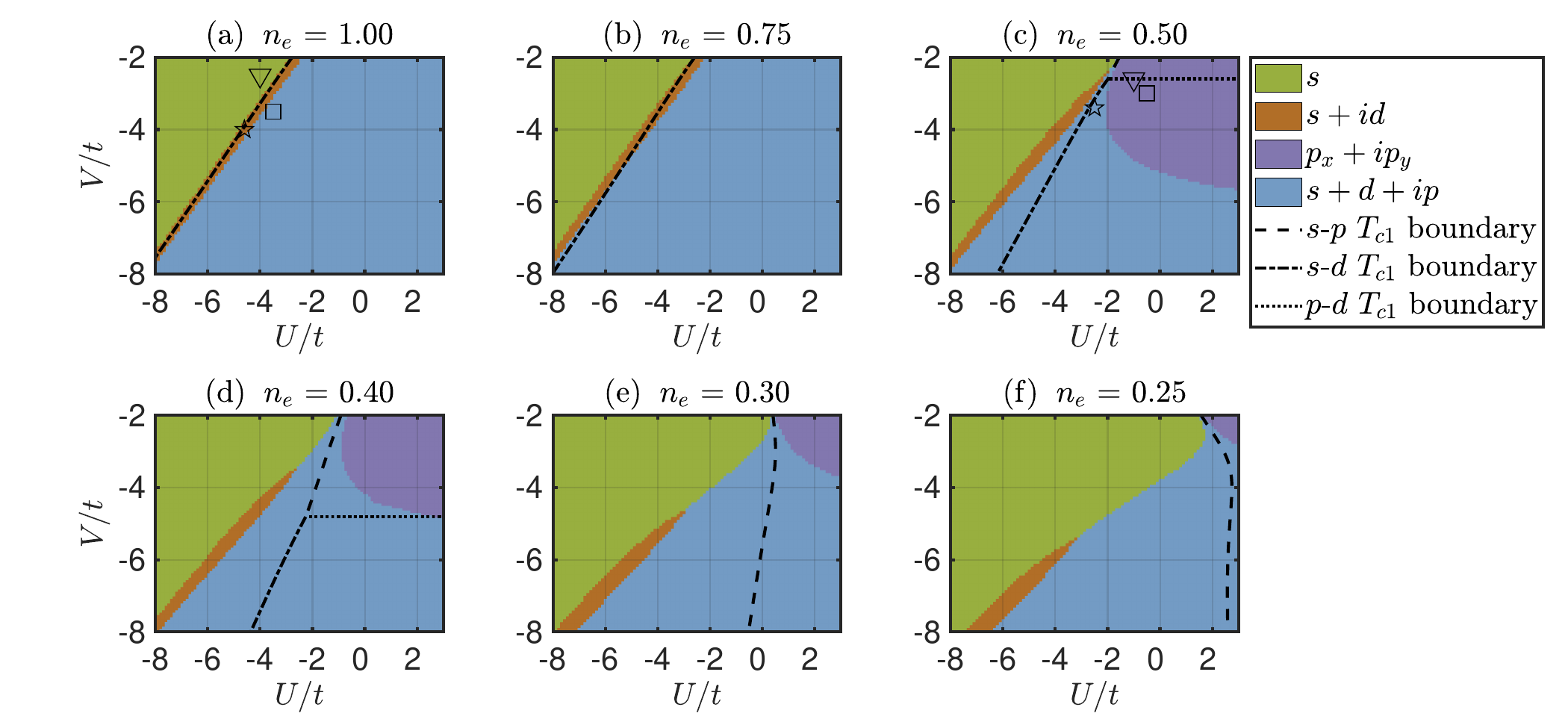}
        \captionsetup{justification=justified, singlelinecheck=false} % Justify multi-line captions
    \caption{\justifying % Ensure justification for the caption
   2D phase diagrams at $T = 0$ for various $n_e$ using $[N_x,\,N_y]$ = $[100,\,100]$ lattice sites. The $\bigtriangledown$, $\square$ and {\small{\FiveStarOpen}} markers on subfigures (a) and (c) correspond to the values of $U$ and $V$ for which the order parameters are plotted in Fig.~\ref{Fig:2Dne100Cooling} and Fig.~\ref{Fig:2Dne050Cooling}.}
    \label{Fig2DBCST0Diagrams}
\end{figure*}

The extended $s$-wave order parameter in Fig.~\ref{Fig:2Dne100Cooling}(b) is barely visible as its magnitude is small in comparison to the other order parameters. Notably, we have verified that the $T_{c}$ for the emergence of the extended $s$-wave and the $p$-wave order parameters from the $d$-wave are coincident. This means that the transition from the $d$-wave phase into the $s+d+ip$ phase does not happen step-wise by going into intermediate $d+ip$ or $s+d$ phases. As the $s$-wave and the $p$-wave order parameters correspond to different irreducible representations of the point group, there is no a priori reason to expect them to appear at the same temperature. However, we have confirmed numerically that this is not a feature peculiar to this point in the phase diagram, but instead appears in a region of $\{T, U, V, n_e\}$ parameter space.

In Fig.~\ref{Fig:2Dne100Cooling}(c), the first superconducting transition occurs at $T_{c1} = 0.931t/k_B$, followed closely by a second transition at $T_{c2} = 0.877t/k_B$, into a mixed-symmetry phase of on-site $s$- and $d$-wave order. Though we do not plot the relative phase of these order parameters, we note that the $\Delta_0$ and $\Delta_{d_{x^2-y^2}}$ order parameters appear with a $\pi/2$ phase difference between them and thus this is an $s+id$ phase. We see that $|\Delta_{d_{x^2-y^2}}|$ shows a kink at this second transition due to the emergence of the $\Delta_0$ order parameter. Like in the 1D case, the on-site $s$- and extended $s$-wave order parameters do not mix at $n_e = 1.00$ though they share the same symmetry. At all other fillings, these two order parameters are coincident. We show this for the chosen points in the $n_e = 0.50$, 2D $T_{c1}$ phase diagrams in Fig.~\ref{Fig:2Dne050Cooling}.

At $T_{c1}$, Fig.~\ref{Fig:2Dne050Cooling}(a) corresponds to a $p$-wave phase, while Fig.~\ref{Fig:2Dne050Cooling}(b) and Fig.~\ref{Fig:2Dne050Cooling}(c) correspond to a $d$-wave phase. In Fig.~\ref{Fig:2Dne050Cooling}(a) we see the superconducting transition occurs at $T_{c1} = 0.236t/k_B$ into the chiral $p_x+ip_y$ phase~\cite{Kallin2015}; note that the $|\Delta_{p_y}|$ curve lies atop the $|\Delta_{p_x}|$ curve. This is identified as $p_x+ip_y$ as both $|\Delta_{p_y}|$ and $|\Delta_{p_x}|$ are both equal in magnitude and appear with a $\pi/2$ phase difference. In this case, we see that there is no second superconducting transition, and this phase persists to $T = 0$.

In Fig.~\ref{Fig:2Dne050Cooling}(b) we see three superconducting transitions: the first occurs at $T_{c1} = 0.413t/k_B$ into a pure $d$-wave phase and the second transition occurs at $T_{c2} = 0.375t/k_B$ into the $s+d+ip$ mixed-symmetry phase with all the order parameters being non-zero. However, the $\Delta_0$ order parameter is much smaller in magnitude than the other order parameters, and is nearly invisible in this subfigure. The third and final transition occurs at $T_{c3} = 0.285t/k_B$ into a pure $p$-wave $p_x+ip_y$ phase. 

Note that the $d$ $\rightarrow$ $s+d+ip$ transition is a second-order (or continuous) phase transition, while the transition from the $s+d+ip$ to the $p_x+ip_y$ phase is a first-order phase transition. As discussed in Appendix C of Reference~\cite{Hutchinson_2020}, these transitions can be understood as bifurcations of the stationary points of the free energy. At the initial superconducting transition at $T_{c1} = 0.413t/k_B$, the free-energy landscape develops three minima corresponding to the three pure symmetry phases --- of these three stationary points, the pure $d$-wave phase remains the global minimum until $T_{c2} = 0.375t/k_B$. At this temperature, the pure $d$-wave stationary point bifurcates further and the mixed-symmetry phase emerges as the global minimum. At $T_{c3} = 0.285t/k_B$ the pure $p$-wave phase minimum becomes the lowest free-energy solution.

In Fig.~\ref{Fig:2Dne050Cooling}(c) the first superconducting transition occurs at $T_{c1} = 0.589t/k_B$, followed closely by a second transition at $T_{c2} = 0.526t/k_B$, into the completely mixed-symmetry $s+d+ip$ phase, which remains the stable phase until $T = 0$. Similar to Fig.~\ref{Fig:2Dne100Cooling}(b) this phase transition does not happen step-wise though an intermediate $d+s$ or $d+ip$ phase. As we are away from half filling, this $s+d+ip$ phase has both on-site $s$- and extended $s$-wave order parameters.

\subsection{2D phase diagrams at $T = 0$}

Having seen how these select points on the phase diagrams evolve upon cooling, we now set $T=0$ and examine the 2D ground-state phase diagrams. In Fig.~\ref{Fig2DBCST0Diagrams} we plot the 2D $T=0$ phase diagrams with the $T_{c1}$ phase boundaries from Fig.~\ref{Fig2DBCSTcDiagrams} as black dashed and/or dotted lines on each $T=0$ phase diagram.  However, due to the larger number of possible order parameter combinations in 2D, the colours in the phase diagrams are organized differently than in the 1D case of Fig.~\ref{Fig1DBCST0Diagrams}. We now aggregate the order parameters by their symmetry --- thus ``$s$'' in the legend now refers to phases which could have just $\Delta_0$ non-zero, just $\Delta_{s^*}$ non-zero, or both $\Delta_0$ and $\Delta_{s^*}$ non-zero.

We see that the chiral $p_x+ip_y$ phase is stable over the largest region for $n_e = 0.50$, and it gets smaller as we reduce the electron density due to encroachment by both the pure $s$ and mixed $s+d+ip$ phases. For all $n_e$, we see an $s+id$ phase as a narrow strip, sandwiched between the pure $s$ and mixed $s+d+ip$ phases. As we reduce the electron density, it only survives at high values of attractive $U$ and $V$.

\subsection{2D density of states}
As in the 1D case, we focus on two representative $U$-$V$ points for the 2D DOS: $U/t = -3.50$ and $V/t = -3.50$ for $n_e = 1.00$, and $U/t = -0.50$ and $V/t = -3.00$ for $n_e = 0.50$. Respectively, these points are the $\square$ and {\small{\FiveStarOpen}} markers in subfigures (a) and (c) of Fig.~\ref{Fig2DBCSTcDiagrams} and Fig.~\ref{Fig2DBCST0Diagrams}. We show the 2D DOS for all the $\bigtriangledown$, $\square$ and {\small{\FiveStarOpen}} markers in Appendix~\ref{App2}. 

\begin{figure}[t]
    \centering
    \includegraphics[width=0.95\linewidth]{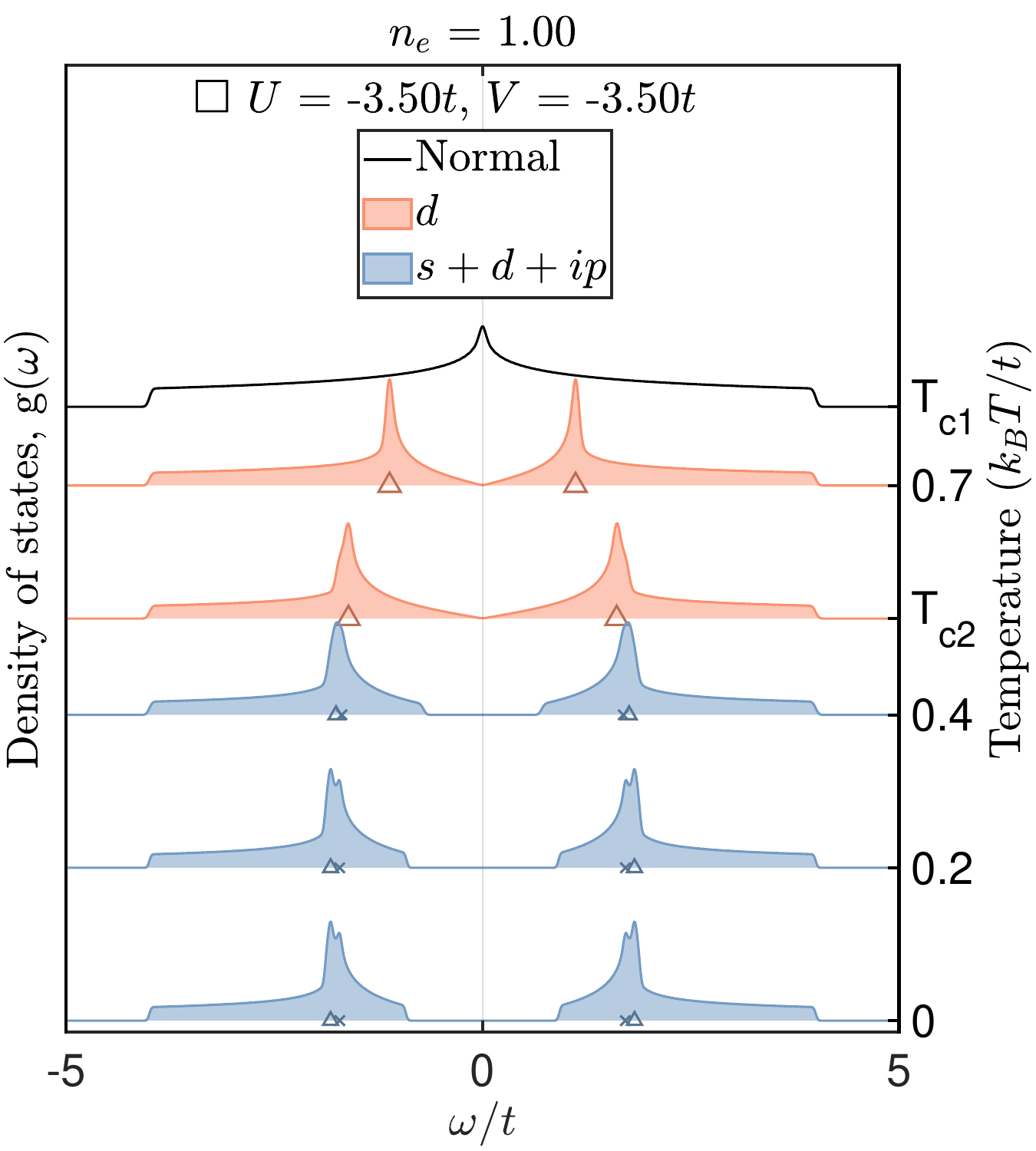}
    \caption{\justifying 2D density of states for $n_e = 1.00$ at $U/t = -3.50$ and $V/t = -3.50$ ($\square$ in subfigure (a) of Fig.~\ref{Fig2DBCSTcDiagrams} and Fig.~\ref{Fig2DBCST0Diagrams}). The transition temperatures are $T_{c1} = 0.803t/k_B$ and $T_{c2} = 0.525t/k_B$, and the colour of each DOS plot corresponds to its phase in its phase diagram. ``Normal'' in the legend refers to the normal state (non-superconducting) 2D density of states. The $\scriptstyle\triangle$ markers indicate the location of the Van Hove singularities originating from the normal state, and the $\times$ markers indicate additional emergent peaks.
    % At $T_{c1}$ the superconducting phase is a pure $d$-wave phase, and undergoes a phase transition to a mixed-symmetry $s+d+ip$ phase at $T_{c2} = 0.525t/k_B$.
    }
    \label{Fig:Waterfall2Dne100DOS}
\end{figure}

In contrast to the 1D case, the normal state 2D DOS has a single logarithmic Van Hove singularity located at $\omega = -\mu$. Thus at half-filling (where $\mu =0$), this singularity is situated where the superconducting gap opens up. In Fig.~\ref{Fig:Waterfall2Dne100DOS} we show how this singularity evolves when the system first enters a pure $d$-wave superconducting phase, before transitioning to a mixed-symmetry $s+d+ip$ phase below $T_{c2}$. In contrast to the other superconducting phases considered in this study, the pure $d$-wave phase is anisotropic and fully closes at $k_x = \pm k_y$. Thus it has a nodal gap, and the DOS increases linearly with $\omega$ at low energy. In 2D, singularities in the DOS arise due to saddle points~ in $E_{\boldsymbol{k}}$; in the $d$-wave phase at $n_e = 1.00$~\cite{Zhou1992}, two logarithmic Van Hove singularities emerge at

\begin{equation}
    \omega_{\scriptstyle\triangle} = \pm  \frac{2t|\Delta_{d_{x^2-y^2}}|}{\sqrt{4t^2+|\Delta_{d_{x^2-y^2}}|^2}}.
\end{equation}

We can interpret these two peaks as the BCS coherence peaks, or as the splitting of the normal state Van Hove singularity by the superconducting gap. For this reason, we indicate the location of these peaks with $\scriptstyle\triangle$ markers in Fig.~\ref{Fig:Waterfall2Dne100DOS}. Cooling below $T_{c2}$ into the mixed-symmetry $s+d+ip$ phase, we see two main features: i) the DOS becomes fully gapped, and ii) the two peaks identified in the $d$-wave phase each split into two peaks. These peaks are located at the minima of $E_{\boldsymbol{k}}$ along the $k_x =0$ and $k_y =0$ lines:

\begin{equation}\label{EqOmegatriadncross}
\begin{aligned}
    \omega_{\scriptstyle\triangle} &= \pm \min(E_{\boldsymbol{k}|_{k_x=0}}),\\
    \omega_{\times} &= \pm \min(E_{\boldsymbol{k}|_{k_y=0}}).
\end{aligned}
\end{equation}

Note that along the $k_x=0$ line, the $\Delta_{p_x}$ contribution to $E_{\boldsymbol{k}}$ disappears, while in the orthogonal direction it does not. Thus this splitting of the DOS peaks occurs \PSY{here} due to the mixed-symmetry nature of the superconducting phase, including both singlet and triplet order parameters. This statement holds true if the saddle point corresponding to $\omega_{\times}$ does not occur at $k_x = 0$ or $k_x=\pm\pi$, which we have checked is not the case here.

\begin{figure}[t]
    \centering
    \includegraphics[width=0.95\linewidth]{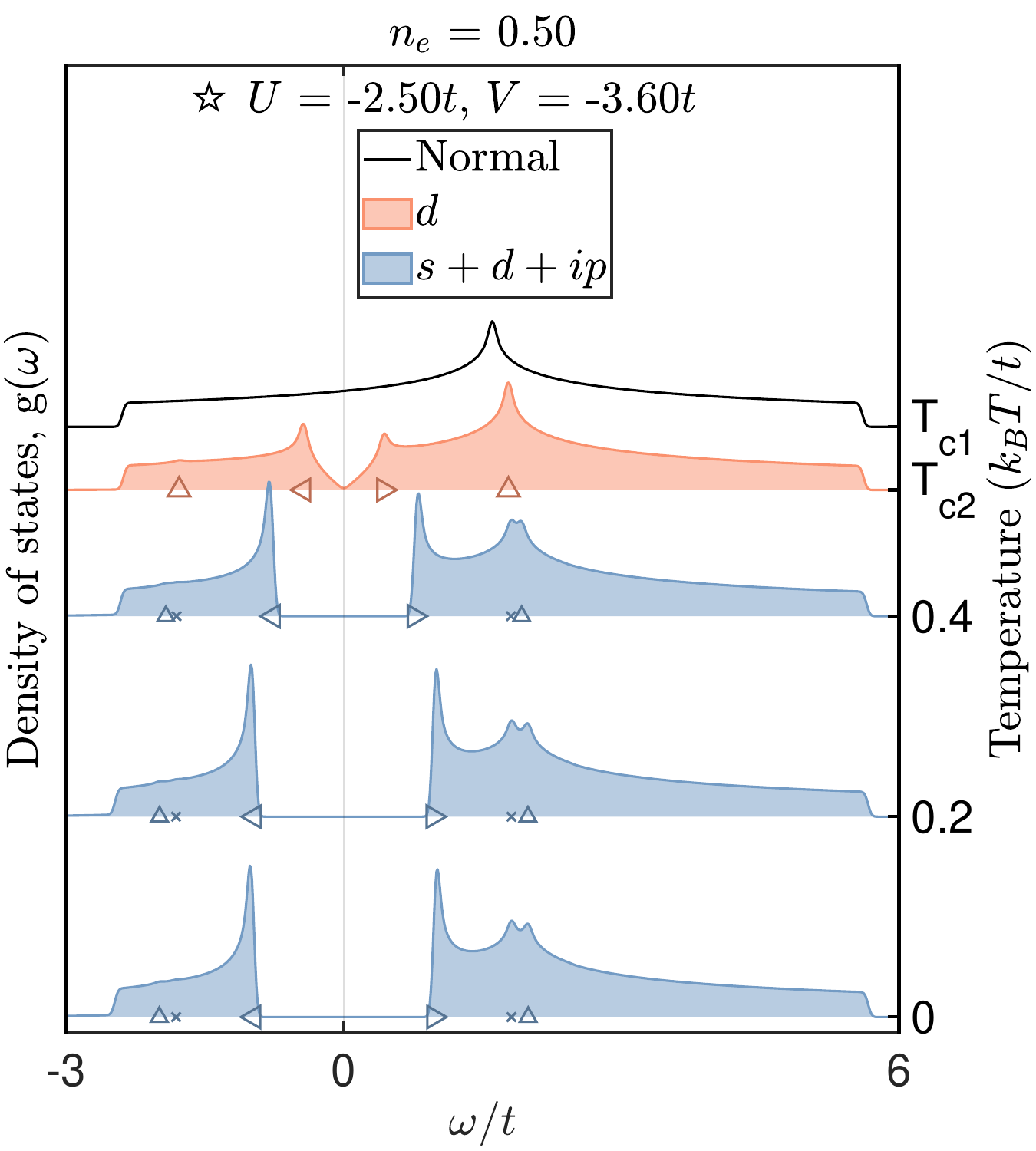}
    \caption{\justifying 2D density of states for $n_e = 0.50$ at $U/t = -2.50$ and $V/t = -3.60$ ({\small{\FiveStarOpen}} in subfigure (c) of Fig.~\ref{Fig2DBCSTcDiagrams} and Fig.~\ref{Fig2DBCST0Diagrams}).  The transition temperatures are $T_{c1} = 0.589t/k_B$ and $T_{c2} = 0.526t/k_B$, and the colour of each DOS plot corresponds to its phase in its phase diagram. ``Normal'' in the legend refers to the normal state (non-superconducting) 2D density of states. The $\scriptstyle\triangle$ markers indicate the location of the Van Hove singularities originating from the normal state, the $\triangleright$ and $\triangleleft$ markers indicate the location of gap-edge DOS peaks, and the $\times$ markers indicate additional emergent peaks.
    % At $T_{c1}$ the superconducting phase is a pure $d$-wave phase, and undergoes a phase transition to a mixed-symmetry $s+d+ip$ phase at $T_{c2} = 0.375t/k_B$, and finally to a chiral $p_x+ip_y$ phase at $T_{c3} = 0.285t/k_B$
    }
    \label{Fig:Waterfall2Dne050DOS_2}
\end{figure}

In Fig.~\ref{Fig:Waterfall2Dne050DOS_2}, we consider the $n_e = 0.50$ case, again with a pure $d$-wave superconducting phase transitioning to a mixed-symmetry $s+d+ip$ phase below $T_{c2}$. In this case, the normal state Van Hove singularity is located away from the gap, and thus we can disentangle its identity from the BCS coherence peaks. Below half-filling, the two lowest energy saddle points of the $d$-wave phase~\cite{Zhou1992} lead to singularities at

\begin{equation}
    \omega_{\triangleright,\triangleleft} = \pm  \frac12 \frac{(4t+\mu)|\Delta_{d_{x^2-y^2}}|}{\sqrt{4t^2+|\Delta_{d_{x^2-y^2}}|^2}}.
\end{equation}

The peak we identify with the original Van Hove singularity appears from an energetically higher saddle point of $E_{\boldsymbol{k}}$, and occurs at

\begin{equation}
    \omega_{\scriptstyle\triangle} = \pm  \sqrt{\mu^2+|\Delta_{d_{x^2-y^2}}|^2}.
\end{equation}

However we note that the peak in the negative frequency band is heavily suppressed, and is hard to pick out in this DOS plot. 

Cooling below $T_{c2}$ into the $s+d+ip$ phase, we again see the splitting of the Van Hove peaks. These occur due to saddle points in $E_{\boldsymbol{k}}$ at $(k_x,\,k_y) =(0,\pm\pi)$ and $(k_x,\,k_y) =(\pm\pi,0)$, which lead to peaks located at

\begin{equation}
\begin{aligned}
    \omega_{\scriptstyle\triangle} &= \pm \sqrt{\mu^2+|\Delta_0+\Delta_{d_{x^2-y^2}}|^2},\\
    \omega_{\times} &= \pm \sqrt{\mu^2+|\Delta_0-\Delta_{d_{x^2-y^2}}|^2}.
\end{aligned}
\end{equation}

Thus the splitting of these peaks originates only from the interplay between the onsite $s$-wave and $d$-wave order parameters, and do not involve the extended $s$-wave or $p$-wave order parameters. The BCS coherence peaks at the gap edge instead correspond to the solutions of Equation~\ref{EqOmegatriadncross}:

\begin{equation}
     \omega_{\triangleright,\triangleleft} = \pm \min(E_{\boldsymbol{k}|_{k_x=0}}) \approx \pm \min(E_{\boldsymbol{k}|_{k_y=0}}).
\end{equation}

Note that the seemingly lone peaks at the gap edge arise from the \textit{approximate} equality of the minima in the above equation, coupled with the Gaussian broadening we apply. We find that in most regions of the zero temperature $\{U, V, n_e\}$ parameter space for the $s+d+ip$ phase, this energy difference is minimal and we obtain a three-peak structure in each frequency band. However, a well-defined four-peak structure does occur for some parameter regions --- we provide an example of this at $n_e = 0.65$ in Fig.~\ref{Fig:2Dne065DOS} of Appendix~\ref{App2}.

To summarize, the 2D DOS in the mixed-symmetry $s+d+ip$ phase shows a distinctive multi-peak structure, emerging from the splitting of the Van Hove/BCS coherence peaks. We observe a two-peak structure at half-filling, and a three-peak (and occasionally four-peak) structure below half-filling. Amongst the superconducting phases studied in this section, this feature is unique to the $s+d+ip$ phase and has its origin in the mix of symmetries of this phase. Three-peak structures in the DOS have been observed in a quasi-2D organic superconductor~\cite{Guterding2016} but with a nodal gap, originating from a mixed-symmetry extended $s$- and $d$-wave phase. Though we do not find this phase to be energetically favourable in our study, it provides impetus for further study of mixed-symmetry phases in different systems. (Note that the $s^*+d$ phase is distinct from the $s+id$ phase which we find, as the $s+id$ is fully gapped --- see Fig.~\ref{Fig:2Dne100DOS} in Appendix~\ref{App2}.)

\section{Discussion}\label{SecIV}
We have investigated the emergence of mixed-symmetry superconductivity in the translationally invariant extended Hubbard model at low temperatures on the 1D lattice and the 2D square lattice. We accomplished this using the BCS method, which involves performing a mean-field approximation in $\bm{k}$-space, and pairing the electrons in a $(\bm{k} \uparrow,-\bm{k} \downarrow)$ state. The on-site interaction, $U$, and the nearest-neighbour interaction, $V$, restrict us to a set of three order parameters $\{\Delta_0,\, \Delta_s^*,\, \Delta_{p}\}$  in 1D, and  a set of five order parameters $\{\Delta_0,\, \Delta_s^*,\, \Delta_{d_{x^2-y^2}},\, \Delta_{p_x},\, \Delta_{p_y}\}$ in 2D. The symmetry of the gap can be classified as pure $s$-wave, $p$-wave or $d$-wave at $T_{c1}$, but a variety of mixed-symmetry phases emerge as stable phases as we cool the system further. At $T=0$, our results show rich phase diagrams of both pure and mixed-symmetry superconducting phases.

On the 1D lattice, we found all three possible combinations of the order parameters were present at $T=0$: the (i) pure $s$, (ii) pure $p$, and (iii) mixed-symmetry $s+ip$ phases. This latter phase is the only possible mixed phase if we assume the superconducting gap to be unitary. Away from half filling, the pure $p$ region shrinks, and the phase diagram is dominated by the pure $s$ and mixed $s+ip$ phases.

On the 2D square lattice, the larger set of order parameters leads to richer phase diagrams at $T=0$. The pure $s$-wave phase is stable for large regions of the $U$-$V$ parameter space at all fillings. The pure $p$-wave phase --- taking the chiral $p_x+ip_y$ form --- is most stable at $n_e = 0.50$ and occupies a large region of parameter space near repulsive $U$. The most stable mixed-symmetry phase is the $s+d+ip$ phase and occupies the largest region of parameter space of all the phases at $n_e = 0.75$. At lower fillings, its stability is encroached upon, first by the pure $p$-wave phase and then the pure $s$-wave phase. The mixed-symmetry $s+id$ phase is also present at all values of $n_e$ we explored; however, it only exists in a narrow region of parameter space near $|U| = |V|$ and disappears from the low $U$ and $V$ region as we lower $n_e$.

Finally, we studied the density of states of these superconducting phases in 1D and 2D. We found that the origin of the peaks in the DOS could be distinguished as either from the Van Hove peaks of the normal state, or the BCS coherence peaks, and that these peaks could split upon transitions into different superconducting phases. In particular, we discovered that the mixed-symmetry $s+d+ip$ phase showcases a distinctive multi-peak structure in the DOS, with its origin being directly related to the interplay of the different symmetry order parameters. Through this simple mean-field model, we have identified signatures of symmetry which could be relevant to experiments. This applies both to real materials, such as the plethora of quasi-2D organic superconductors which have been shown to permit mixed-symmetry superconductivity~\cite{Powell_2006,Wosnitza2007}, and to quantum simulators using ultracold Fermi gases in optical lattices~\cite{Tarruell2018,Takahashi2022}. To unravel these possible symmetry transitions within the superconducting state, it is important to perform measurements over the entire range of superconducting temperatures.

\begin{acknowledgments}
 \PSY{We thank Majid Kheirkhah for helpful correspondence regarding the reason for the $\pi/2$ phase shift in the mixed-symmetry order parameter.} PSY acknowledges financial support by the Alberta Innovates Graduate Student Scholarship Program and the European Research Council through the Advanced Grant DyMETEr (10.3030/101054500). This work was supported in part by the Natural Sciences and Engineering Research Council of Canada (NSERC) and by a MIF from the Province of Alberta. 

 We wish to acknowledge the memory of Jan Zaanen with this manuscript. His animated discussion of a multitude of topics at conferences and accompanying meals will always be remembered. His enthusiasm for discovery and for ``new ways of thinking about things'' remain an inspiration for those who knew him.

\end{acknowledgments}
\FloatBarrier
\vskip0.2in

\bibliography{biblio}

\begin{thebibliography}{50}%
\makeatletter
\providecommand \@ifxundefined [1]{%
 \@ifx{#1\undefined}
}%
\providecommand \@ifnum [1]{%
 \ifnum #1\expandafter \@firstoftwo
 \else \expandafter \@secondoftwo
 \fi
}%
\providecommand \@ifx [1]{%
 \ifx #1\expandafter \@firstoftwo
 \else \expandafter \@secondoftwo
 \fi
}%
\providecommand \natexlab [1]{#1}%
\providecommand \enquote  [1]{``#1''}%
\providecommand \bibnamefont  [1]{#1}%
\providecommand \bibfnamefont [1]{#1}%
\providecommand \citenamefont [1]{#1}%
\providecommand \href@noop [0]{\@secondoftwo}%
\providecommand \href [0]{\begingroup \@sanitize@url \@href}%
\providecommand \@href[1]{\@@startlink{#1}\@@href}%
\providecommand \@@href[1]{\endgroup#1\@@endlink}%
\providecommand \@sanitize@url [0]{\catcode `\\12\catcode `\$12\catcode `\&12\catcode `\#12\catcode `\^12\catcode `\_12\catcode `\%12\relax}%
\providecommand \@@startlink[1]{}%
\providecommand \@@endlink[0]{}%
\providecommand \url  [0]{\begingroup\@sanitize@url \@url }%
\providecommand \@url [1]{\endgroup\@href {#1}{\urlprefix }}%
\providecommand \urlprefix  [0]{URL }%
\providecommand \Eprint [0]{\href }%
\providecommand \doibase [0]{https://doi.org/}%
\providecommand \selectlanguage [0]{\@gobble}%
\providecommand \bibinfo  [0]{\@secondoftwo}%
\providecommand \bibfield  [0]{\@secondoftwo}%
\providecommand \translation [1]{[#1]}%
\providecommand \BibitemOpen [0]{}%
\providecommand \bibitemStop [0]{}%
\providecommand \bibitemNoStop [0]{.\EOS\space}%
\providecommand \EOS [0]{\spacefactor3000\relax}%
\providecommand \BibitemShut  [1]{\csname bibitem#1\endcsname}%
\let\auto@bib@innerbib\@empty
%</preamble>
\bibitem [{\citenamefont {Norman}(2011)}]{Norman2011}%
  \BibitemOpen
  \bibfield  {author} {\bibinfo {author} {\bibfnamefont {M.~R.}\ \bibnamefont {Norman}},\ }\bibfield  {title} {\bibinfo {title} {The challenge of unconventional superconductivity},\ }\href {https://doi.org/10.1126/science.1200181} {\bibfield  {journal} {\bibinfo  {journal} {Science}\ }\textbf {\bibinfo {volume} {332}},\ \bibinfo {pages} {196} (\bibinfo {year} {2011})}\BibitemShut {NoStop}%
\bibitem [{\citenamefont {Fischer}\ \emph {et~al.}(2007)\citenamefont {Fischer}, \citenamefont {Kugler}, \citenamefont {Maggio-Aprile}, \citenamefont {Berthod},\ and\ \citenamefont {Renner}}]{fischer07}%
  \BibitemOpen
  \bibfield  {author} {\bibinfo {author} {\bibfnamefont {O.}~\bibnamefont {Fischer}}, \bibinfo {author} {\bibfnamefont {M.}~\bibnamefont {Kugler}}, \bibinfo {author} {\bibfnamefont {I.}~\bibnamefont {Maggio-Aprile}}, \bibinfo {author} {\bibfnamefont {C.}~\bibnamefont {Berthod}},\ and\ \bibinfo {author} {\bibfnamefont {C.}~\bibnamefont {Renner}},\ }\bibfield  {title} {\bibinfo {title} {Scanning tunneling spectroscopy of high-temperature superconductors},\ }\href {https://doi.org/10.1103/RevModPhys.79.353} {\bibfield  {journal} {\bibinfo  {journal} {Rev. Mod. Phys.}\ }\textbf {\bibinfo {volume} {79}},\ \bibinfo {pages} {353} (\bibinfo {year} {2007})}\BibitemShut {NoStop}%
\bibitem [{\citenamefont {Sobota}\ \emph {et~al.}(2021)\citenamefont {Sobota}, \citenamefont {He},\ and\ \citenamefont {Shen}}]{ARPESreview}%
  \BibitemOpen
  \bibfield  {author} {\bibinfo {author} {\bibfnamefont {J.~A.}\ \bibnamefont {Sobota}}, \bibinfo {author} {\bibfnamefont {Y.}~\bibnamefont {He}},\ and\ \bibinfo {author} {\bibfnamefont {Z.-X.}\ \bibnamefont {Shen}},\ }\bibfield  {title} {\bibinfo {title} {{Angle-resolved photoemission studies of quantum materials}},\ }\href {https://doi.org/10.1103/RevModPhys.93.025006} {\bibfield  {journal} {\bibinfo  {journal} {Rev. Mod. Phys.}\ }\textbf {\bibinfo {volume} {93}},\ \bibinfo {pages} {025006} (\bibinfo {year} {2021})}\BibitemShut {NoStop}%
\bibitem [{\citenamefont {Van~Harlingen}(1995)}]{vanHarlingen1995}%
  \BibitemOpen
  \bibfield  {author} {\bibinfo {author} {\bibfnamefont {D.~J.}\ \bibnamefont {Van~Harlingen}},\ }\bibfield  {title} {\bibinfo {title} {Phase-sensitive tests of the symmetry of the pairing state in the high-temperature superconductors---evidence for ${d}_{{x}^{2}\ensuremath{-}{y}^{2}}$ symmetry},\ }\href {https://doi.org/10.1103/RevModPhys.67.515} {\bibfield  {journal} {\bibinfo  {journal} {Rev. Mod. Phys.}\ }\textbf {\bibinfo {volume} {67}},\ \bibinfo {pages} {515} (\bibinfo {year} {1995})}\BibitemShut {NoStop}%
\bibitem [{\citenamefont {Bardeen}\ \emph {et~al.}(1957)\citenamefont {Bardeen}, \citenamefont {Cooper},\ and\ \citenamefont {Schrieffer}}]{BCS1957}%
  \BibitemOpen
  \bibfield  {author} {\bibinfo {author} {\bibfnamefont {J.}~\bibnamefont {Bardeen}}, \bibinfo {author} {\bibfnamefont {L.~N.}\ \bibnamefont {Cooper}},\ and\ \bibinfo {author} {\bibfnamefont {J.~R.}\ \bibnamefont {Schrieffer}},\ }\bibfield  {title} {\bibinfo {title} {{Microscopic Theory of Superconductivity}},\ }\href {https://doi.org/10.1103/PhysRev.106.162} {\bibfield  {journal} {\bibinfo  {journal} {Phys. Rev.}\ }\textbf {\bibinfo {volume} {106}},\ \bibinfo {pages} {162} (\bibinfo {year} {1957})}\BibitemShut {NoStop}%
\bibitem [{\citenamefont {Pitaevskii}(1959)}]{Pitaevskii1959}%
  \BibitemOpen
  \bibfield  {author} {\bibinfo {author} {\bibfnamefont {L.~P.}\ \bibnamefont {Pitaevskii}},\ }\bibfield  {title} {\bibinfo {title} {{On superfluidity of liquid {He}$-3$}},\ }\bibfield  {journal} {\bibinfo  {journal} {Zhur. Eksptl'. i Teoret. Fiz.}\ }\textbf {\bibinfo {volume} {37}},\ \href {https://www.osti.gov/biblio/4146818} {} (\bibinfo {year} {1959})\BibitemShut {NoStop}%
\bibitem [{\citenamefont {Brueckner}\ \emph {et~al.}(1960)\citenamefont {Brueckner}, \citenamefont {Soda}, \citenamefont {Anderson},\ and\ \citenamefont {Morel}}]{brueckner1960level}%
  \BibitemOpen
  \bibfield  {author} {\bibinfo {author} {\bibfnamefont {K.~A.}\ \bibnamefont {Brueckner}}, \bibinfo {author} {\bibfnamefont {T.}~\bibnamefont {Soda}}, \bibinfo {author} {\bibfnamefont {P.~W.}\ \bibnamefont {Anderson}},\ and\ \bibinfo {author} {\bibfnamefont {P.}~\bibnamefont {Morel}},\ }\bibfield  {title} {\bibinfo {title} {Level structure of nuclear matter and liquid {He}$^3$},\ }\href {https://doi.org/10.1103/PhysRev.118.1442} {\bibfield  {journal} {\bibinfo  {journal} {Phys. Rev.}\ }\textbf {\bibinfo {volume} {118}},\ \bibinfo {pages} {1442} (\bibinfo {year} {1960})}\BibitemShut {NoStop}%
\bibitem [{\citenamefont {Emery}\ and\ \citenamefont {Sessler}(1960)}]{emery1960possible}%
  \BibitemOpen
  \bibfield  {author} {\bibinfo {author} {\bibfnamefont {V.~J.}\ \bibnamefont {Emery}}\ and\ \bibinfo {author} {\bibfnamefont {A.~M.}\ \bibnamefont {Sessler}},\ }\bibfield  {title} {\bibinfo {title} {Possible phase transition in liquid {He}$^3$},\ }\href {https://doi.org/10.1103/PhysRev.119.43} {\bibfield  {journal} {\bibinfo  {journal} {Phys. Rev.}\ }\textbf {\bibinfo {volume} {119}},\ \bibinfo {pages} {43} (\bibinfo {year} {1960})}\BibitemShut {NoStop}%
\bibitem [{\citenamefont {Thouless}(1960)}]{thouless1960perturbation}%
  \BibitemOpen
  \bibfield  {author} {\bibinfo {author} {\bibfnamefont {D.~J.}\ \bibnamefont {Thouless}},\ }\bibfield  {title} {\bibinfo {title} {Perturbation theory in statistical mechanics and the theory of superconductivity},\ }\href {https://doi.org/10.1016/0003-4916(60)90122-6} {\bibfield  {journal} {\bibinfo  {journal} {Ann. Phys.}\ }\textbf {\bibinfo {volume} {10}},\ \bibinfo {pages} {553} (\bibinfo {year} {1960})}\BibitemShut {NoStop}%
\bibitem [{\citenamefont {Anderson}\ and\ \citenamefont {Morel}(1961)}]{anderson1961generalized}%
  \BibitemOpen
  \bibfield  {author} {\bibinfo {author} {\bibfnamefont {P.~W.}\ \bibnamefont {Anderson}}\ and\ \bibinfo {author} {\bibfnamefont {P.}~\bibnamefont {Morel}},\ }\bibfield  {title} {\bibinfo {title} {{Generalized Bardeen-Cooper-Schrieffer states and the proposed low-temperature phase of liquid {He}$^3$}},\ }\href {https://doi.org/10.1103/PhysRev.123.1911} {\bibfield  {journal} {\bibinfo  {journal} {Phys. Rev.}\ }\textbf {\bibinfo {volume} {123}},\ \bibinfo {pages} {1911} (\bibinfo {year} {1961})}\BibitemShut {NoStop}%
\bibitem [{\citenamefont {Balian}\ and\ \citenamefont {Werthamer}(1963)}]{balian1963superconductivity}%
  \BibitemOpen
  \bibfield  {author} {\bibinfo {author} {\bibfnamefont {R.}~\bibnamefont {Balian}}\ and\ \bibinfo {author} {\bibfnamefont {N.~R.}\ \bibnamefont {Werthamer}},\ }\bibfield  {title} {\bibinfo {title} {Superconductivity with pairs in a relative $p$-wave},\ }\href {https://doi.org/10.1103/PhysRev.131.1553} {\bibfield  {journal} {\bibinfo  {journal} {Phys. Rev.}\ }\textbf {\bibinfo {volume} {131}},\ \bibinfo {pages} {1553} (\bibinfo {year} {1963})}\BibitemShut {NoStop}%
\bibitem [{\citenamefont {Anderson}(2011)}]{Anderson2011}%
  \BibitemOpen
  \bibfield  {author} {\bibinfo {author} {\bibfnamefont {P.~W.}\ \bibnamefont {Anderson}},\ }\bibfield  {title} {\bibinfo {title} {{He-3, Pierre Morel and Me{\textemdash}Early Work on Anisotropic Superfluidity}},\ }\href {https://doi.org/10.1007/s10909-011-0368-7} {\bibfield  {journal} {\bibinfo  {journal} {J. Low Temp. Phys.}\ }\textbf {\bibinfo {volume} {164}},\ \bibinfo {pages} {119} (\bibinfo {year} {2011})}\BibitemShut {NoStop}%
\bibitem [{\citenamefont {Sessler}(2011)}]{Sessler2011}%
  \BibitemOpen
  \bibfield  {author} {\bibinfo {author} {\bibfnamefont {A.~M.}\ \bibnamefont {Sessler}},\ }\bibfield  {title} {\bibinfo {title} {{Early Thoughts on the Superfluidity of Helium-3}},\ }\href {https://doi.org/10.1007/s10909-011-0369-6} {\bibfield  {journal} {\bibinfo  {journal} {J. Low Temp. Phys.}\ }\textbf {\bibinfo {volume} {164}},\ \bibinfo {pages} {126–139} (\bibinfo {year} {2011})}\BibitemShut {NoStop}%
\bibitem [{\citenamefont {Lee}\ and\ \citenamefont {Leggett}(2011)}]{Lee2011}%
  \BibitemOpen
  \bibfield  {author} {\bibinfo {author} {\bibfnamefont {D.~M.}\ \bibnamefont {Lee}}\ and\ \bibinfo {author} {\bibfnamefont {A.~J.}\ \bibnamefont {Leggett}},\ }\bibfield  {title} {\bibinfo {title} {{Superfluid $^3$He—the Early Days}},\ }\href {https://doi.org/10.1007/s10909-011-0375-8} {\bibfield  {journal} {\bibinfo  {journal} {J. Low Temp. Phys.}\ }\textbf {\bibinfo {volume} {164}},\ \bibinfo {pages} {140–172} (\bibinfo {year} {2011})}\BibitemShut {NoStop}%
\bibitem [{\citenamefont {Pitaevskii}(2011)}]{Pitaevskii2011}%
  \BibitemOpen
  \bibfield  {author} {\bibinfo {author} {\bibfnamefont {L.~P.}\ \bibnamefont {Pitaevskii}},\ }\bibfield  {title} {\bibinfo {title} {{“On the Superfluidity of Liquid He3”. Author’s Recollections}},\ }\href {https://doi.org/10.1007/s10909-011-0367-8} {\bibfield  {journal} {\bibinfo  {journal} {J. Low Temp. Phys.}\ }\textbf {\bibinfo {volume} {164}},\ \bibinfo {pages} {173–177} (\bibinfo {year} {2011})}\BibitemShut {NoStop}%
\bibitem [{\citenamefont {Annett}(1990)}]{Annett1990}%
  \BibitemOpen
  \bibfield  {author} {\bibinfo {author} {\bibfnamefont {J.~F.}\ \bibnamefont {Annett}},\ }\bibfield  {title} {\bibinfo {title} {Symmetry of the order parameter for high-temperature superconductivity},\ }\href {https://doi.org/10.1080/00018739000101481} {\bibfield  {journal} {\bibinfo  {journal} {Adv. Phys.}\ }\textbf {\bibinfo {volume} {39}},\ \bibinfo {pages} {83–126} (\bibinfo {year} {1990})}\BibitemShut {NoStop}%
\bibitem [{\citenamefont {Sigrist}\ and\ \citenamefont {Ueda}(1991)}]{SigristUeda1991}%
  \BibitemOpen
  \bibfield  {author} {\bibinfo {author} {\bibfnamefont {M.}~\bibnamefont {Sigrist}}\ and\ \bibinfo {author} {\bibfnamefont {K.}~\bibnamefont {Ueda}},\ }\bibfield  {title} {\bibinfo {title} {{Phenomenological theory of unconventional superconductivity}},\ }\href {https://doi.org/10.1103/RevModPhys.63.239} {\bibfield  {journal} {\bibinfo  {journal} {Rev. Mod. Phys.}\ }\textbf {\bibinfo {volume} {63}},\ \bibinfo {pages} {239} (\bibinfo {year} {1991})}\BibitemShut {NoStop}%
\bibitem [{\citenamefont {Sigrist}(2005)}]{Sigrist2005}%
  \BibitemOpen
  \bibfield  {author} {\bibinfo {author} {\bibfnamefont {M.}~\bibnamefont {Sigrist}},\ }\bibfield  {title} {\bibinfo {title} {Introduction to unconventional superconductivity},\ }in\ \href {https://doi.org/10.1063/1.2080350} {\emph {\bibinfo {booktitle} {AIP Conference Proceedings}}},\ Vol.\ \bibinfo {volume} {789}\ (\bibinfo  {publisher} {AIP},\ \bibinfo {year} {2005})\ p.\ \bibinfo {pages} {165–243}\BibitemShut {NoStop}%
\bibitem [{\citenamefont {Normand}\ \emph {et~al.}(1994)\citenamefont {Normand}, \citenamefont {Kohno},\ and\ \citenamefont {Fukuyama}}]{NORMAND1994}%
  \BibitemOpen
  \bibfield  {author} {\bibinfo {author} {\bibfnamefont {B.}~\bibnamefont {Normand}}, \bibinfo {author} {\bibfnamefont {H.}~\bibnamefont {Kohno}},\ and\ \bibinfo {author} {\bibfnamefont {H.}~\bibnamefont {Fukuyama}},\ }\bibfield  {title} {\bibinfo {title} {{Properties of a mixed-symmetry superconductor}},\ }\href {https://doi.org/https://doi.org/10.1016/0921-4534(94)92359-0} {\bibfield  {journal} {\bibinfo  {journal} {Physica C: Superconductivity}\ }\textbf {\bibinfo {volume} {235-240}},\ \bibinfo {pages} {2275} (\bibinfo {year} {1994})}\BibitemShut {NoStop}%
\bibitem [{\citenamefont {O’Donovan}\ and\ \citenamefont {Carbotte}(1995)}]{ODonovan1995}%
  \BibitemOpen
  \bibfield  {author} {\bibinfo {author} {\bibfnamefont {C.}~\bibnamefont {O’Donovan}}\ and\ \bibinfo {author} {\bibfnamefont {J.}~\bibnamefont {Carbotte}},\ }\bibfield  {title} {\bibinfo {title} {{Mixed order parameter symmetry in the BCS model}},\ }\href {https://doi.org/10.1016/0921-4534(95)00451-3} {\bibfield  {journal} {\bibinfo  {journal} {Physica C: Superconductivity}\ }\textbf {\bibinfo {volume} {252}},\ \bibinfo {pages} {87–99} (\bibinfo {year} {1995})}\BibitemShut {NoStop}%
\bibitem [{\citenamefont {Musaelian}\ \emph {et~al.}(1996)\citenamefont {Musaelian}, \citenamefont {Betouras}, \citenamefont {Chubukov},\ and\ \citenamefont {Joynt}}]{Musaelian1996}%
  \BibitemOpen
  \bibfield  {author} {\bibinfo {author} {\bibfnamefont {K.~A.}\ \bibnamefont {Musaelian}}, \bibinfo {author} {\bibfnamefont {J.}~\bibnamefont {Betouras}}, \bibinfo {author} {\bibfnamefont {A.~V.}\ \bibnamefont {Chubukov}},\ and\ \bibinfo {author} {\bibfnamefont {R.}~\bibnamefont {Joynt}},\ }\bibfield  {title} {\bibinfo {title} {{Mixed-symmetry superconductivity in two-dimensional Fermi liquids}},\ }\href {https://doi.org/10.1103/PhysRevB.53.3598} {\bibfield  {journal} {\bibinfo  {journal} {Phys. Rev. B}\ }\textbf {\bibinfo {volume} {53}},\ \bibinfo {pages} {3598} (\bibinfo {year} {1996})}\BibitemShut {NoStop}%
\bibitem [{\citenamefont {Betouras}(1997)}]{betouras1997}%
  \BibitemOpen
  \bibfield  {author} {\bibinfo {author} {\bibfnamefont {J.}~\bibnamefont {Betouras}},\ }\href {https://books.google.at/books?id=cYNpAAAAMAAJ} {\emph {\bibinfo {title} {{Superconductivity with Mixed-symmetry Gap Function}}}}\ (\bibinfo  {publisher} {University of Wisconsin--Madison},\ \bibinfo {year} {1997})\BibitemShut {NoStop}%
\bibitem [{\citenamefont {Mitra}\ \emph {et~al.}(1998)\citenamefont {Mitra}, \citenamefont {Ghosh},\ and\ \citenamefont {Behera}}]{Mitra1998}%
  \BibitemOpen
  \bibfield  {author} {\bibinfo {author} {\bibfnamefont {M.}~\bibnamefont {Mitra}}, \bibinfo {author} {\bibfnamefont {H.}~\bibnamefont {Ghosh}},\ and\ \bibinfo {author} {\bibfnamefont {S.}~\bibnamefont {Behera}},\ }\bibfield  {title} {\bibinfo {title} {{Effect of electron correlation on superconducting pairing symmetry}},\ }\href {https://doi.org/10.1007/s100510050260} {\bibfield  {journal} {\bibinfo  {journal} {The European Physical Journal B}\ }\textbf {\bibinfo {volume} {2}},\ \bibinfo {pages} {371–380} (\bibinfo {year} {1998})}\BibitemShut {NoStop}%
\bibitem [{\citenamefont {Angilella}\ and\ \citenamefont {Pucci}(2001)}]{ANGILELLA2001}%
  \BibitemOpen
  \bibfield  {author} {\bibinfo {author} {\bibfnamefont {G.}~\bibnamefont {Angilella}}\ and\ \bibinfo {author} {\bibfnamefont {R.}~\bibnamefont {Pucci}},\ }\bibfield  {title} {\bibinfo {title} {{Mixed-symmetry solutions of the BCS gap equation}},\ }\href {https://doi.org/https://doi.org/10.1016/S0362-546X(01)00471-0} {\bibfield  {journal} {\bibinfo  {journal} {Nonlinear Analysis: Theory, Methods \& Applications}\ }\textbf {\bibinfo {volume} {47}},\ \bibinfo {pages} {3537} (\bibinfo {year} {2001})},\ \bibinfo {note} {{Proceedings of the Third World Congress of Nonlinear Analysts}}\BibitemShut {NoStop}%
\bibitem [{\citenamefont {Basu}(2006)}]{BASU2006}%
  \BibitemOpen
  \bibfield  {author} {\bibinfo {author} {\bibfnamefont {S.}~\bibnamefont {Basu}},\ }\bibfield  {title} {\bibinfo {title} {{Gap functions in anisotropic superconductors}},\ }\href {https://doi.org/https://doi.org/10.1016/j.physb.2006.01.150} {\bibfield  {journal} {\bibinfo  {journal} {Physica B: Condensed Matter}\ }\textbf {\bibinfo {volume} {378-380}},\ \bibinfo {pages} {430} (\bibinfo {year} {2006})},\ \bibinfo {note} {{Proceedings of the International Conference on Strongly Correlated Electron Systems}}\BibitemShut {NoStop}%
\bibitem [{\citenamefont {Medvedev}(2012)}]{Medvedev2012}%
  \BibitemOpen
  \bibfield  {author} {\bibinfo {author} {\bibfnamefont {M.~V.}\ \bibnamefont {Medvedev}},\ }\bibfield  {title} {\bibinfo {title} {{Superconducting states with a mixed symmetry of order parameters in the model of a two-dimensional Fermi liquid}},\ }\href {https://doi.org/10.1134/s0031918x12020093} {\bibfield  {journal} {\bibinfo  {journal} {The Physics of Metals and Metallography}\ }\textbf {\bibinfo {volume} {113}},\ \bibinfo {pages} {117–128} (\bibinfo {year} {2012})}\BibitemShut {NoStop}%
\bibitem [{\citenamefont {Timirgazin}\ \emph {et~al.}(2019)\citenamefont {Timirgazin}, \citenamefont {Gilmutdinov},\ and\ \citenamefont {Arzhnikov}}]{Timirgazin2019}%
  \BibitemOpen
  \bibfield  {author} {\bibinfo {author} {\bibfnamefont {M.}~\bibnamefont {Timirgazin}}, \bibinfo {author} {\bibfnamefont {V.}~\bibnamefont {Gilmutdinov}},\ and\ \bibinfo {author} {\bibfnamefont {A.}~\bibnamefont {Arzhnikov}},\ }\bibfield  {title} {\bibinfo {title} {{Phase diagrams of singlet superconducting states with mixed symmetry}},\ }\href {https://doi.org/https://doi.org/10.1016/j.physc.2018.12.003} {\bibfield  {journal} {\bibinfo  {journal} {Physica C: Superconductivity and its Applications}\ }\textbf {\bibinfo {volume} {557}},\ \bibinfo {pages} {7} (\bibinfo {year} {2019})}\BibitemShut {NoStop}%
\bibitem [{\citenamefont {Akbar}\ \emph {et~al.}(2024)\citenamefont {Akbar}, \citenamefont {Biborski}, \citenamefont {Rademaker},\ and\ \citenamefont {Zegrodnik}}]{Akbar2024}%
  \BibitemOpen
  \bibfield  {author} {\bibinfo {author} {\bibfnamefont {W.}~\bibnamefont {Akbar}}, \bibinfo {author} {\bibfnamefont {A.}~\bibnamefont {Biborski}}, \bibinfo {author} {\bibfnamefont {L.}~\bibnamefont {Rademaker}},\ and\ \bibinfo {author} {\bibfnamefont {M.}~\bibnamefont {Zegrodnik}},\ }\bibfield  {title} {\bibinfo {title} {{Topological superconductivity with mixed singlet-triplet pairing in moir\'e transition metal dichalcogenide bilayers}},\ }\href {https://doi.org/10.1103/PhysRevB.110.064516} {\bibfield  {journal} {\bibinfo  {journal} {Phys. Rev. B}\ }\textbf {\bibinfo {volume} {110}},\ \bibinfo {pages} {064516} (\bibinfo {year} {2024})}\BibitemShut {NoStop}%
\bibitem [{\citenamefont {Sutradhar}\ \emph {et~al.}(2024)\citenamefont {Sutradhar}, \citenamefont {Ruhman},\ and\ \citenamefont {Klein}}]{Sutradhar2024}%
  \BibitemOpen
  \bibfield  {author} {\bibinfo {author} {\bibfnamefont {J.}~\bibnamefont {Sutradhar}}, \bibinfo {author} {\bibfnamefont {J.}~\bibnamefont {Ruhman}},\ and\ \bibinfo {author} {\bibfnamefont {A.}~\bibnamefont {Klein}},\ }\bibfield  {title} {\bibinfo {title} {{Singlet, triplet, and mixed all-to-all pairing states emerging from incoherent fermions}},\ }\href {https://doi.org/10.1103/PhysRevResearch.6.L042036} {\bibfield  {journal} {\bibinfo  {journal} {Phys. Rev. Res.}\ }\textbf {\bibinfo {volume} {6}},\ \bibinfo {pages} {L042036} (\bibinfo {year} {2024})}\BibitemShut {NoStop}%
\bibitem [{\citenamefont {Kheirkhah}\ \emph {et~al.}(2020)\citenamefont {Kheirkhah}, \citenamefont {Yan}, \citenamefont {Nagai},\ and\ \citenamefont {Marsiglio}}]{Kheirkhah2020}%
  \BibitemOpen
  \bibfield  {author} {\bibinfo {author} {\bibfnamefont {M.}~\bibnamefont {Kheirkhah}}, \bibinfo {author} {\bibfnamefont {Z.}~\bibnamefont {Yan}}, \bibinfo {author} {\bibfnamefont {Y.}~\bibnamefont {Nagai}},\ and\ \bibinfo {author} {\bibfnamefont {F.}~\bibnamefont {Marsiglio}},\ }\bibfield  {title} {\bibinfo {title} {{First- and Second-Order Topological Superconductivity and Temperature-Driven Topological Phase Transitions in the Extended Hubbard Model with Spin-Orbit Coupling}},\ }\href {https://doi.org/10.1103/PhysRevLett.125.017001} {\bibfield  {journal} {\bibinfo  {journal} {Phys. Rev. Lett.}\ }\textbf {\bibinfo {volume} {125}},\ \bibinfo {pages} {017001} (\bibinfo {year} {2020})}\BibitemShut {NoStop}%
\bibitem [{\citenamefont {S{\"o}rensen}\ \emph {et~al.}(1991)\citenamefont {S{\"o}rensen}, \citenamefont {Schneider},\ and\ \citenamefont {Frick}}]{Sorensen1991}%
  \BibitemOpen
  \bibfield  {author} {\bibinfo {author} {\bibfnamefont {M.~P.}\ \bibnamefont {S{\"o}rensen}}, \bibinfo {author} {\bibfnamefont {T.}~\bibnamefont {Schneider}},\ and\ \bibinfo {author} {\bibfnamefont {M.}~\bibnamefont {Frick}},\ }\bibinfo {title} {{Microscopic Aspects of Nonlinearity in Condensed Matter, Nonlinear Properties of the BCS Gap Equation}}\ (\bibinfo  {publisher} {Springer US},\ \bibinfo {address} {Boston, MA},\ \bibinfo {year} {1991})\ pp.\ \bibinfo {pages} {315--327}\BibitemShut {NoStop}%
\bibitem [{\citenamefont {Kuboki}(2001)}]{Kuboki2001}%
  \BibitemOpen
  \bibfield  {author} {\bibinfo {author} {\bibfnamefont {K.}~\bibnamefont {Kuboki}},\ }\bibfield  {title} {\bibinfo {title} {{Effect of Band Structure on the Symmetry of Superconducting States}},\ }\href {https://doi.org/10.1143/jpsj.70.2698} {\bibfield  {journal} {\bibinfo  {journal} {Journal of the Physical Society of Japan}\ }\textbf {\bibinfo {volume} {70}},\ \bibinfo {pages} {2698–2702} (\bibinfo {year} {2001})}\BibitemShut {NoStop}%
\bibitem [{\citenamefont {Nayak}\ and\ \citenamefont {Kumar}(2018)}]{Nayak_2018}%
  \BibitemOpen
  \bibfield  {author} {\bibinfo {author} {\bibfnamefont {S.}~\bibnamefont {Nayak}}\ and\ \bibinfo {author} {\bibfnamefont {S.}~\bibnamefont {Kumar}},\ }\bibfield  {title} {\bibinfo {title} {{Exotic superconducting states in the extended attractive {{Hubbard}} model}},\ }\href {https://doi.org/10.1088/1361-648x/aaaefe} {\bibfield  {journal} {\bibinfo  {journal} {J. Phys. Condens. Matter}\ }\textbf {\bibinfo {volume} {30}},\ \bibinfo {pages} {135601} (\bibinfo {year} {2018})}\BibitemShut {NoStop}%
\bibitem [{\citenamefont {Hutchinson}\ and\ \citenamefont {Marsiglio}(2020)}]{Hutchinson_2020}%
  \BibitemOpen
  \bibfield  {author} {\bibinfo {author} {\bibfnamefont {J.}~\bibnamefont {Hutchinson}}\ and\ \bibinfo {author} {\bibfnamefont {F.}~\bibnamefont {Marsiglio}},\ }\bibfield  {title} {\bibinfo {title} {{Mixed temperature-dependent order parameters in the extended {{Hubbard}} model}},\ }\href {https://doi.org/10.1088/1361-648x/abc801} {\bibfield  {journal} {\bibinfo  {journal} {J. Phys. Condens. Matter}\ }\textbf {\bibinfo {volume} {33}},\ \bibinfo {pages} {065603} (\bibinfo {year} {2020})}\BibitemShut {NoStop}%
\bibitem [{\citenamefont {Richardson}(1963)}]{richardson63}%
  \BibitemOpen
  \bibfield  {author} {\bibinfo {author} {\bibfnamefont {R.~W.}\ \bibnamefont {Richardson}},\ }\bibfield  {title} {\bibinfo {title} {{A restricted class of exact eigenstates of the pairing-force Hamiltonian}},\ }\href {https://doi.org/https://doi.org/10.1016/0031-9163(63)90259-2} {\bibfield  {journal} {\bibinfo  {journal} {Phys. Lett.}\ }\textbf {\bibinfo {volume} {3}},\ \bibinfo {pages} {277} (\bibinfo {year} {1963})}\BibitemShut {NoStop}%
\bibitem [{\citenamefont {Richardson}\ and\ \citenamefont {Sherman}(1964)}]{richardson64}%
  \BibitemOpen
  \bibfield  {author} {\bibinfo {author} {\bibfnamefont {R.~W.}\ \bibnamefont {Richardson}}\ and\ \bibinfo {author} {\bibfnamefont {N.}~\bibnamefont {Sherman}},\ }\bibfield  {title} {\bibinfo {title} {{Exact eigenstates of the pairing-force Hamiltonian}},\ }\href {https://doi.org/https://doi.org/10.1016/0029-5582(64)90687-X} {\bibfield  {journal} {\bibinfo  {journal} {Nucl. Phys.}\ }\textbf {\bibinfo {volume} {52}},\ \bibinfo {pages} {221} (\bibinfo {year} {1964})}\BibitemShut {NoStop}%
\bibitem [{\citenamefont {Richardson}(1977)}]{richardson77}%
  \BibitemOpen
  \bibfield  {author} {\bibinfo {author} {\bibfnamefont {R.~W.}\ \bibnamefont {Richardson}},\ }\bibfield  {title} {\bibinfo {title} {Pairing in the limit of a large number of particles},\ }\href {https://doi.org/10.1063/1.523493} {\bibfield  {journal} {\bibinfo  {journal} {J. Math. Phys.}\ }\textbf {\bibinfo {volume} {18}},\ \bibinfo {pages} {1802} (\bibinfo {year} {1977})}\BibitemShut {NoStop}%
\bibitem [{\citenamefont {Combescot}\ \emph {et~al.}(2013)\citenamefont {Combescot}, \citenamefont {Pogosov},\ and\ \citenamefont {Betbeder-Matibet}}]{combescot13}%
  \BibitemOpen
  \bibfield  {author} {\bibinfo {author} {\bibfnamefont {M.}~\bibnamefont {Combescot}}, \bibinfo {author} {\bibfnamefont {W.}~\bibnamefont {Pogosov}},\ and\ \bibinfo {author} {\bibfnamefont {O.}~\bibnamefont {Betbeder-Matibet}},\ }\bibfield  {title} {\bibinfo {title} {{BCS ansatz for superconductivity in the light of the Bogoliubov approach and the Richardson–Gaudin exact wave function}},\ }\href {https://doi.org/https://doi.org/10.1016/j.physc.2012.10.011} {\bibfield  {journal} {\bibinfo  {journal} {Physica C}\ }\textbf {\bibinfo {volume} {485}},\ \bibinfo {pages} {47} (\bibinfo {year} {2013})}\BibitemShut {NoStop}%
\bibitem [{\citenamefont {Giaever}\ \emph {et~al.}(1962)\citenamefont {Giaever}, \citenamefont {Hart},\ and\ \citenamefont {Megerle}}]{Giaever1962}%
  \BibitemOpen
  \bibfield  {author} {\bibinfo {author} {\bibfnamefont {I.}~\bibnamefont {Giaever}}, \bibinfo {author} {\bibfnamefont {H.~R.}\ \bibnamefont {Hart}},\ and\ \bibinfo {author} {\bibfnamefont {K.}~\bibnamefont {Megerle}},\ }\bibfield  {title} {\bibinfo {title} {{Tunneling into Superconductors at Temperatures below 1\ifmmode^\circ\else\textdegree\fi{}K}},\ }\href {https://doi.org/10.1103/PhysRev.126.941} {\bibfield  {journal} {\bibinfo  {journal} {Phys. Rev.}\ }\textbf {\bibinfo {volume} {126}},\ \bibinfo {pages} {941} (\bibinfo {year} {1962})}\BibitemShut {NoStop}%
\bibitem [{\citenamefont {Tsuei}\ and\ \citenamefont {Kirtley}(2000)}]{Tsuei2000}%
  \BibitemOpen
  \bibfield  {author} {\bibinfo {author} {\bibfnamefont {C.~C.}\ \bibnamefont {Tsuei}}\ and\ \bibinfo {author} {\bibfnamefont {J.~R.}\ \bibnamefont {Kirtley}},\ }\bibfield  {title} {\bibinfo {title} {Pairing symmetry in cuprate superconductors},\ }\href {https://doi.org/10.1103/RevModPhys.72.969} {\bibfield  {journal} {\bibinfo  {journal} {Rev. Mod. Phys.}\ }\textbf {\bibinfo {volume} {72}},\ \bibinfo {pages} {969} (\bibinfo {year} {2000})}\BibitemShut {NoStop}%
\bibitem [{\citenamefont {Hashimoto}\ \emph {et~al.}(2014)\citenamefont {Hashimoto}, \citenamefont {Vishik}, \citenamefont {He}, \citenamefont {Devereaux},\ and\ \citenamefont {Shen}}]{Hashimoto2014}%
  \BibitemOpen
  \bibfield  {author} {\bibinfo {author} {\bibfnamefont {M.}~\bibnamefont {Hashimoto}}, \bibinfo {author} {\bibfnamefont {I.~M.}\ \bibnamefont {Vishik}}, \bibinfo {author} {\bibfnamefont {R.-H.}\ \bibnamefont {He}}, \bibinfo {author} {\bibfnamefont {T.~P.}\ \bibnamefont {Devereaux}},\ and\ \bibinfo {author} {\bibfnamefont {Z.-X.}\ \bibnamefont {Shen}},\ }\bibfield  {title} {\bibinfo {title} {{Energy gaps in high-transition-temperature cuprate superconductors}},\ }\href {https://doi.org/10.1038/nphys3009} {\bibfield  {journal} {\bibinfo  {journal} {Nature Physics}\ }\textbf {\bibinfo {volume} {10}},\ \bibinfo {pages} {483–495} (\bibinfo {year} {2014})}\BibitemShut {NoStop}%
\bibitem [{\citenamefont {Ramires}(2022)}]{Ramires_2022}%
  \BibitemOpen
  \bibfield  {author} {\bibinfo {author} {\bibfnamefont {A.}~\bibnamefont {Ramires}},\ }\bibfield  {title} {\bibinfo {title} {{Nonunitary superconductivity in complex quantum materials}},\ }\href {https://doi.org/10.1088/1361-648X/ac6d3a} {\bibfield  {journal} {\bibinfo  {journal} {J. Phys. Condens. Matter}\ }\textbf {\bibinfo {volume} {34}},\ \bibinfo {pages} {304001} (\bibinfo {year} {2022})}\BibitemShut {NoStop}%
\bibitem [{\citenamefont {{P. Senarath Yapa}}(2024)}]{PramodhPhDThesis}%
  \BibitemOpen
  \bibfield  {author} {\bibinfo {author} {\bibnamefont {{P. Senarath Yapa}}},\ }\emph {\bibinfo {title} {{Unconventional Cooper Pairing and Confinement: Studies of lattice superconductivity and superfluid helium-3 enclosed by surfaces}}},\ \href@noop {} {Ph.D. thesis},\ \bibinfo  {school} {University of Alberta} (\bibinfo {year} {2024})\BibitemShut {NoStop}%
\bibitem [{\citenamefont {Kallin}\ and\ \citenamefont {Berlinsky}(2016)}]{Kallin2015}%
  \BibitemOpen
  \bibfield  {author} {\bibinfo {author} {\bibfnamefont {C.}~\bibnamefont {Kallin}}\ and\ \bibinfo {author} {\bibfnamefont {J.}~\bibnamefont {Berlinsky}},\ }\bibfield  {title} {\bibinfo {title} {{Chiral superconductors}},\ }\href {https://doi.org/10.1088/0034-4885/79/5/054502} {\bibfield  {journal} {\bibinfo  {journal} {Rep. Prog. Phys.}\ }\textbf {\bibinfo {volume} {79}},\ \bibinfo {pages} {054502} (\bibinfo {year} {2016})}\BibitemShut {NoStop}%
\bibitem [{\citenamefont {Zhou}\ and\ \citenamefont {Schulz}(1992)}]{Zhou1992}%
  \BibitemOpen
  \bibfield  {author} {\bibinfo {author} {\bibfnamefont {C.}~\bibnamefont {Zhou}}\ and\ \bibinfo {author} {\bibfnamefont {H.~J.}\ \bibnamefont {Schulz}},\ }\bibfield  {title} {\bibinfo {title} {Density of states and tunneling spectra in two-dimensional $d$-wave superconductors},\ }\href {https://doi.org/10.1103/PhysRevB.45.7397} {\bibfield  {journal} {\bibinfo  {journal} {Phys. Rev. B}\ }\textbf {\bibinfo {volume} {45}},\ \bibinfo {pages} {7397} (\bibinfo {year} {1992})}\BibitemShut {NoStop}%
\bibitem [{\citenamefont {Guterding}\ \emph {et~al.}(2016)\citenamefont {Guterding}, \citenamefont {Diehl}, \citenamefont {Altmeyer}, \citenamefont {Methfessel}, \citenamefont {Tutsch}, \citenamefont {Schubert}, \citenamefont {Lang}, \citenamefont {M\"uller}, \citenamefont {Huth}, \citenamefont {Jeschke}, \citenamefont {Valent\'{\i}}, \citenamefont {Jourdan},\ and\ \citenamefont {Elmers}}]{Guterding2016}%
  \BibitemOpen
  \bibfield  {author} {\bibinfo {author} {\bibfnamefont {D.}~\bibnamefont {Guterding}}, \bibinfo {author} {\bibfnamefont {S.}~\bibnamefont {Diehl}}, \bibinfo {author} {\bibfnamefont {M.}~\bibnamefont {Altmeyer}}, \bibinfo {author} {\bibfnamefont {T.}~\bibnamefont {Methfessel}}, \bibinfo {author} {\bibfnamefont {U.}~\bibnamefont {Tutsch}}, \bibinfo {author} {\bibfnamefont {H.}~\bibnamefont {Schubert}}, \bibinfo {author} {\bibfnamefont {M.}~\bibnamefont {Lang}}, \bibinfo {author} {\bibfnamefont {J.}~\bibnamefont {M\"uller}}, \bibinfo {author} {\bibfnamefont {M.}~\bibnamefont {Huth}}, \bibinfo {author} {\bibfnamefont {H.~O.}\ \bibnamefont {Jeschke}}, \bibinfo {author} {\bibfnamefont {R.}~\bibnamefont {Valent\'{\i}}}, \bibinfo {author} {\bibfnamefont {M.}~\bibnamefont {Jourdan}},\ and\ \bibinfo {author} {\bibfnamefont {H.-J.}\ \bibnamefont {Elmers}},\ }\bibfield  {title} {\bibinfo {title} {Evidence for eight-node mixed-symmetry superconductivity in a correlated organic metal},\ }\href
  {https://doi.org/10.1103/PhysRevLett.116.237001} {\bibfield  {journal} {\bibinfo  {journal} {Phys. Rev. Lett.}\ }\textbf {\bibinfo {volume} {116}},\ \bibinfo {pages} {237001} (\bibinfo {year} {2016})}\BibitemShut {NoStop}%
\bibitem [{\citenamefont {Powell}(2006)}]{Powell_2006}%
  \BibitemOpen
  \bibfield  {author} {\bibinfo {author} {\bibfnamefont {B.~J.}\ \bibnamefont {Powell}},\ }\bibfield  {title} {\bibinfo {title} {Mixed order parameters, accidental nodes and broken time reversal symmetry in organic superconductors: a group theoretical analysis},\ }\href {https://doi.org/10.1088/0953-8984/18/46/L01} {\bibfield  {journal} {\bibinfo  {journal} {J. Phys. Condens. Matter}\ }\textbf {\bibinfo {volume} {18}},\ \bibinfo {pages} {L575} (\bibinfo {year} {2006})}\BibitemShut {NoStop}%
\bibitem [{\citenamefont {Wosnitza}(2007)}]{Wosnitza2007}%
  \BibitemOpen
  \bibfield  {author} {\bibinfo {author} {\bibfnamefont {J.}~\bibnamefont {Wosnitza}},\ }\bibfield  {title} {\bibinfo {title} {Quasi-two-dimensional organic superconductors},\ }\href {https://doi.org/10.1007/s10909-006-9282-9} {\bibfield  {journal} {\bibinfo  {journal} {J. Low Temp. Phys.}\ }\textbf {\bibinfo {volume} {146}},\ \bibinfo {pages} {641–667} (\bibinfo {year} {2007})}\BibitemShut {NoStop}%
\bibitem [{\citenamefont {Tarruell}\ and\ \citenamefont {Sanchez-Palencia}(2018)}]{Tarruell2018}%
  \BibitemOpen
  \bibfield  {author} {\bibinfo {author} {\bibfnamefont {L.}~\bibnamefont {Tarruell}}\ and\ \bibinfo {author} {\bibfnamefont {L.}~\bibnamefont {Sanchez-Palencia}},\ }\bibfield  {title} {\bibinfo {title} {{Quantum simulation of the Hubbard model with ultracold fermions in optical lattices}},\ }\href {https://doi.org/10.1016/j.crhy.2018.10.013} {\bibfield  {journal} {\bibinfo  {journal} {C. R. Phys.}\ }\textbf {\bibinfo {volume} {19}},\ \bibinfo {pages} {365–393} (\bibinfo {year} {2018})}\BibitemShut {NoStop}%
\bibitem [{\citenamefont {Takahashi}(2022)}]{Takahashi2022}%
  \BibitemOpen
  \bibfield  {author} {\bibinfo {author} {\bibfnamefont {Y.}~\bibnamefont {Takahashi}},\ }\bibfield  {title} {\bibinfo {title} {Quantum simulation of quantum many-body systems with ultracold two-electron atoms in an optical lattice},\ }\href {https://doi.org/10.2183/pjab.98.010} {\bibfield  {journal} {\bibinfo  {journal} {Proc. Jpn. Acad. Ser. B}\ }\textbf {\bibinfo {volume} {98}},\ \bibinfo {pages} {141–160} (\bibinfo {year} {2022})}\BibitemShut {NoStop}%
\end{thebibliography}%

\clearpage
\appendix

\section{1D Density of States}\label{App1}

\begin{figure}[b]
    \centering
    \includegraphics[width=0.77\linewidth]{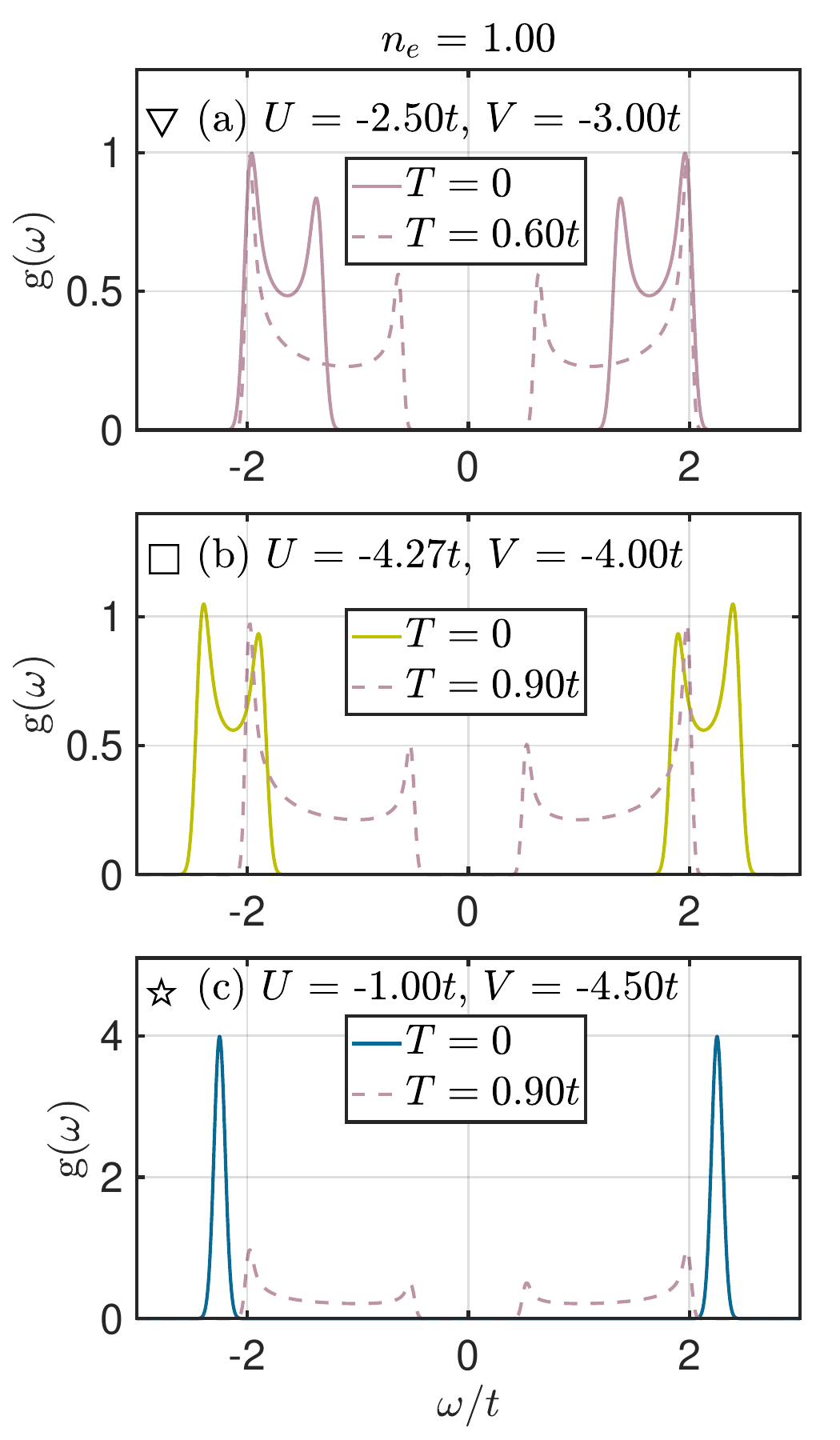}
    \caption{\justifying 1D density of states for $n_e = 1.00$ at $T = 0$ compared with that at $T \sim T_{c1}$. The colour of each line corresponds to its phase in its respective phase diagram. At $T_{c1}$ (a)-(c) all correspond to a pure $p$-wave phase. At $T =0$, the phase (a) remains a pure $p$-wave phase, (b) is a $p$- and on-site $s$-wave phase, and (c) is a $p$-  and extended $s$-wave phase. Fig.~\ref{Fig:Waterfall1Dne100DOS} in the main text corresponds to (b).}
    \label{Fig:1Dne100DOS}
\end{figure}

In this appendix, we show the 1D DOS for all the $U$-$V$ points indicated by the $\bigtriangledown$, $\square$ and {\small{\FiveStarOpen}} markers of Fig.~\ref{Fig1DBCSTcDiagrams} and Fig.~\ref{Fig1DBCST0Diagrams}. In Fig.~\ref{Fig:1Dne100DOS}, we compare the phases at $n_e = 1.00$ by overlaying the DOS for the phase just below $T_{c1}$ and the phase at $T=0$. We can identify the Van Hove and BCS coherence peaks as in the main text, with no additional peaks emerging even at low temperature. Note that the DOS for the $\Delta_{s^*}+i\Delta_p$ phase at $T=0$ in Fig.~\ref{Fig:1Dne100DOS}(c) appears as a single peak due to the two peaks being situated close together and merging due to the Gaussian broadening.

\begin{figure}[b]
    \centering
    \includegraphics[width=0.77\linewidth]{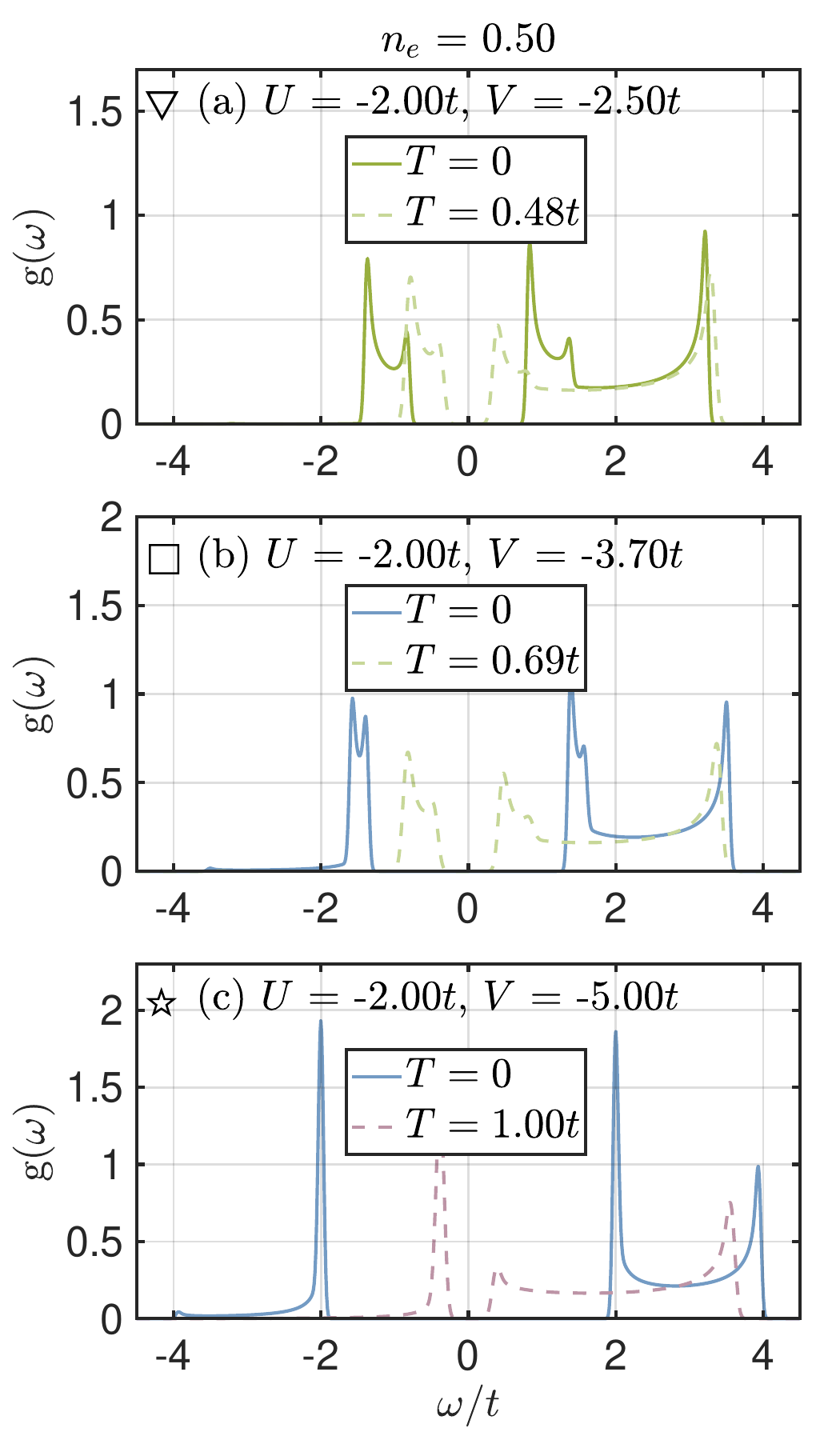}
    \caption{\justifying 1D density of states for $n_e = 0.50$ at $T = 0$ compared with that at $T \sim T_c$. The colour of each line corresponds to its phase in its respective phase diagram. At $T_{c1}$ (a) and (b) correspond to a pure $s$-wave phase, while (c) is a $p$-wave phase. At $T =0$, the phase (a) remains a pure $s$-wave phase of both on-site $s$- and extended $s$-wave order, and both (b) and (c) are mixed-symmetry phases of all three order parameters. Fig.~\ref{Fig:Waterfall1Dne050DOS} in the main text corresponds to (b).}
    \label{Fig:1Dne050DOS}
\end{figure}

 In Fig.~\ref{Fig:1Dne050DOS}, we compare the phases at $n_e = 0.50$ in a similar manner. In these cases, we can identify the additional peak in the positive frequency band in all but Fig.~\ref{Fig:1Dne050DOS}(c). For the $T=0$ mixed-symmetry phase in this subfigure, the peak is present but washed out due to the Gaussian broadening. However for the $p$-wave phase, the peak is not present due to the fact that the additional peak only emerges due to the interplay of different order parameters; as the $p$-wave phase in 1D is only composed of $\Delta_p$, it only shows the standard Van Hove and BCS coherence peaks. Note that the `soft shoulder'' in the negative frequency band is clearly visible in the $T=0$ mixed-symmetry phases of Fig.~\ref{Fig:1Dne050DOS}(b)-(c). The small peak in the DOS at the band edge of negative frequency band (at $\omega = -\omega^+_\triangle$ as defined in the main text) is also clearly visible in these plots.

\section{2D Density of States} \label{App2}

\begin{figure}[t]
    \centering
    \includegraphics[width=0.77\linewidth]{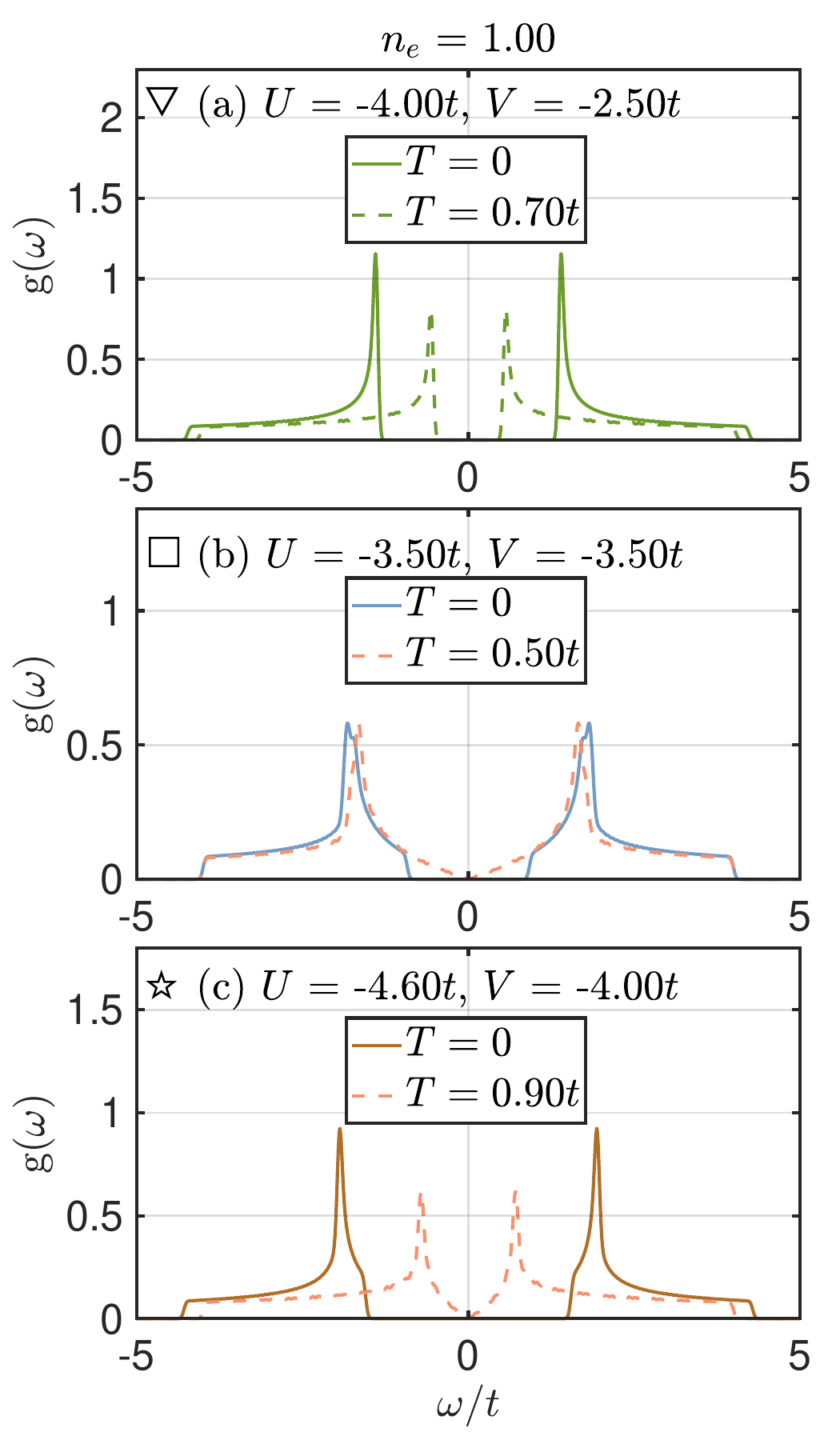}
    \caption{\justifying 2D density of states for $n_e = 1.00$.  The colour of each line corresponds to its phase in its respective phase diagram. At $T_{c1}$ (a) corresponds to a pure $s$-wave phase, and (b) and (c) correspond to pure $d$-wave phases. At $T =0$, the phase (a) remains a pure on-site $s$-wave phase, (b) is a mixed-symmetry $s+d+ip$ phase, and (c) is a mixed-symmetry $s+id$ phase.  Fig.~\ref{Fig:Waterfall2Dne100DOS} in the main text corresponds to (b).}
    \label{Fig:2Dne100DOS}
\end{figure}

In this appendix, we show the 2D DOS for all the $U$-$V$ points indicated by the $\bigtriangledown$, $\square$ and {\small{\FiveStarOpen}} markers of Fig.~\ref{Fig2DBCSTcDiagrams} and Fig.~\ref{Fig2DBCST0Diagrams}. As in the 1D case, in Fig.~\ref{Fig:2Dne100DOS} we compare the phases at $n_e = 1.00$ by overlaying the DOS for the phase just below $T_{c1}$ and the phase at $T=0$. Note that Fig.~\ref{Fig:2Dne100DOS}(c) shows the DOS for the $s+id$ phase at $T=0$. Though this is a mixed-symmetry phase, it does not show any splitting of the DOS peaks due to the relative phase of $\pi/2$ between the $s$-wave and $d$-wave order parameters. We also note that this phase is fully gapped.

\begin{figure}[t]
    \centering
    \includegraphics[width=0.77\linewidth]{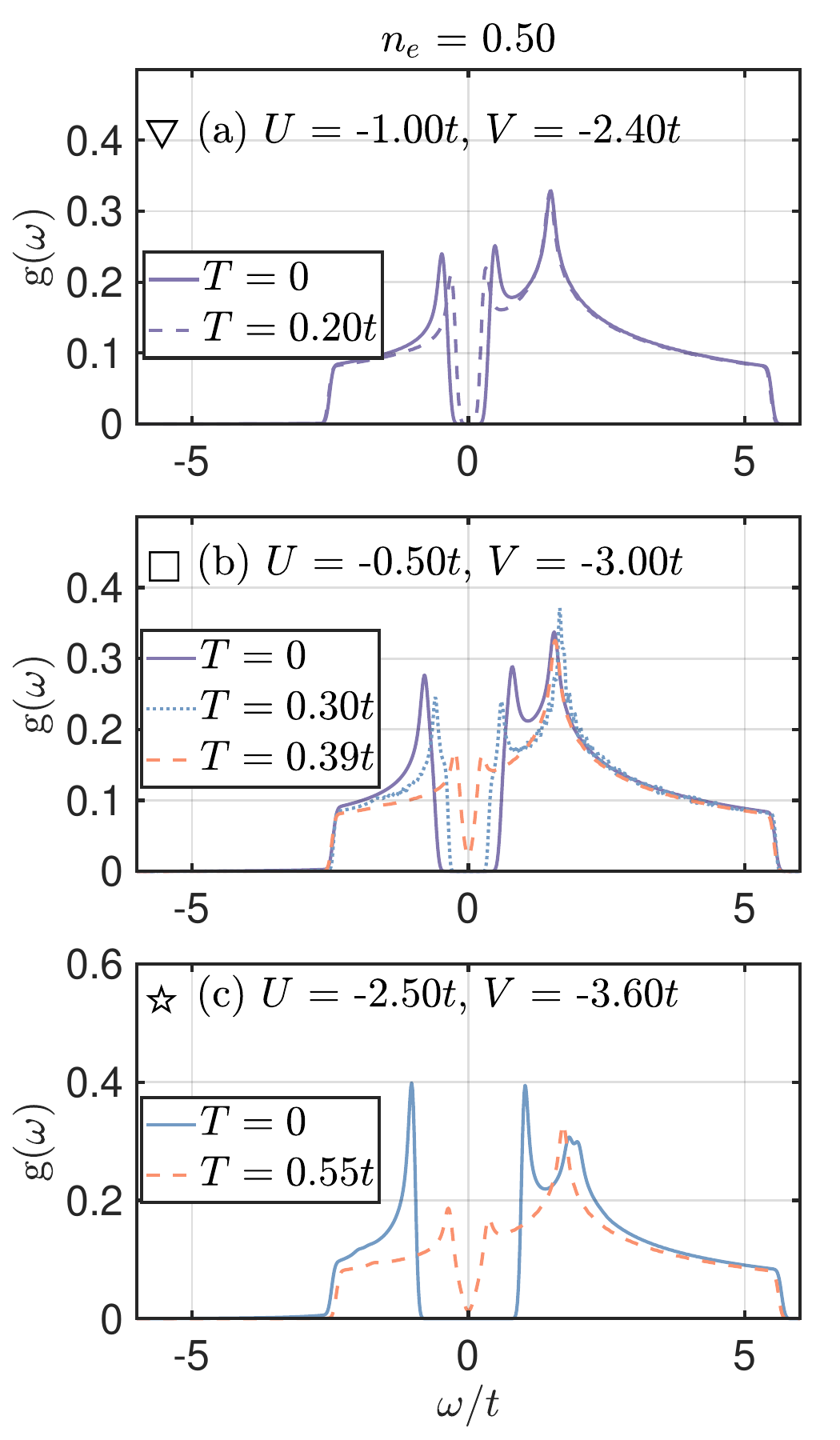}
    \caption{\justifying 2D density of states for $n_e = 0.50$. The colour of each line corresponds to its phase in its respective phase diagram. At $T_{c1}$ (a) corresponds to a pure $p$-wave phase, and (b) and (c) correspond to pure $d$-wave phases. At $T =0$, both (a) and (b) are chiral $p_x+ip_y$ phases, and (c) is a mixed-symmetry $s+d+ip$ phase. Fig.~\ref{Fig:Waterfall2Dne050DOS_2} in the main text corresponds to (c).}
    \label{Fig:2Dne050DOS}
\end{figure}

 In Fig.~\ref{Fig:2Dne050DOS}, we compare the phases at $n_e = 0.50$ in a similar manner but note that Fig.~\ref{Fig:2Dne050DOS}(b) has three DOS plots due to the three superconducting transitions which occur at this $U$-$V$ point. As the $p_x+ip_y$ phases at $T=0$ in Fig.~\ref{Fig:2Dne050DOS}(a)-(b) are pure $p$-wave, their DOS do not contain any additional peaks.

\begin{figure}[b]
    \centering
    \includegraphics[width=0.77\linewidth]{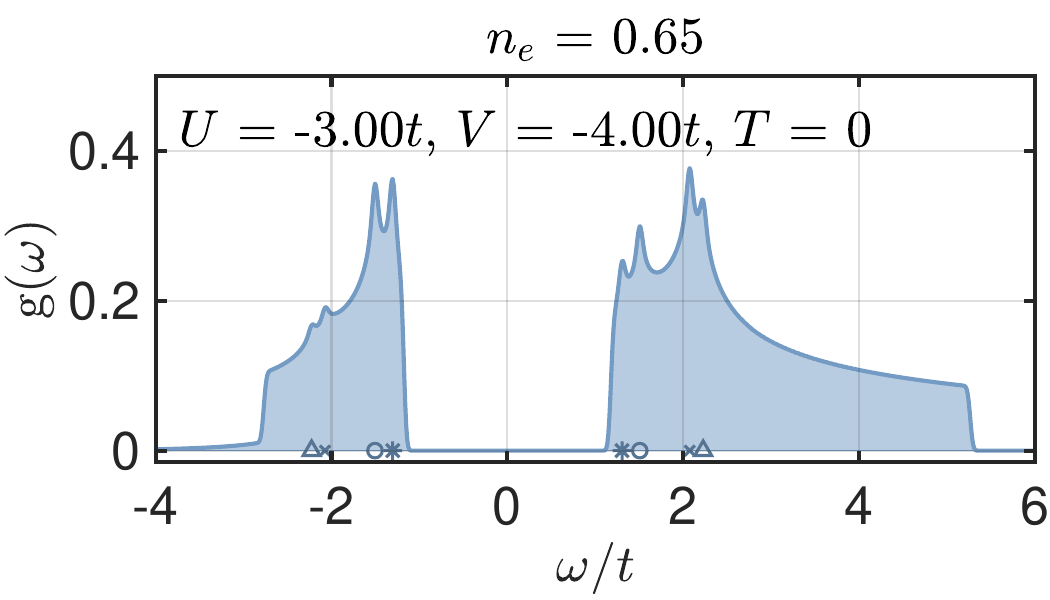}
    \caption{\justifying 2D density of states for $n_e = 0.65$ for the mixed-symmetry $s+d+ip$ phase, showing a four-peak structure in the DOS.}
    \label{Fig:2Dne065DOS}
\end{figure}

 And finally, in Fig.~\ref{Fig:2Dne065DOS} we show an example four-peak DOS for the $s+d+ip$ phase at $n_e = 0.65$ and $T=0$. These peaks occur at:

 \begin{equation}
\begin{aligned}
    \omega_{\scriptstyle\triangle} &= \pm \sqrt{\mu^2+|\Delta_0+\Delta_{d_{x^2-y^2}}|^2},\\
    \omega_{\times} &= \pm \sqrt{\mu^2+|\Delta_0-\Delta_{d_{x^2-y^2}}|^2},\\
    \omega_{\circ} &= \pm \min(E_{\boldsymbol{k}|_{k_x=0}}),\\
    \omega_{*} &= \pm \min(E_{\boldsymbol{k}|_{k_y=0}}).
\end{aligned}
\end{equation}

\FloatBarrier

\end{document}